\pdfoutput=1
\documentclass[reprint,nofootinbib,aps,prl]{revtex4-2}

\usepackage{graphicx}
\usepackage{dcolumn}
\usepackage{booktabs}
\usepackage{bm}
\usepackage[hidelinks]{hyperref}
\usepackage{autobreak}
\usepackage{xcolor}
\usepackage{physics}
\usepackage{amsmath, amssymb} 
\usepackage{mathtools}
\usepackage{ragged2e}
\usepackage{soul} 
\usepackage{orcidlink}

\usepackage{siunitx}
\DeclareSIUnit \parsec {pc}

\usepackage{tikz,lipsum,lmodern}
\usepackage[T1]{fontenc}
\usepackage[most]{tcolorbox}
\usepackage{lmodern}
\usepackage{wrapfig}

\tcbset{colback=green!5,colframe=green!35!black,fonttitle=\bfseries}

\newif\ifdraft \drafttrue
\ifdraft

  \newcommand{\NK}[1]{{\color{teal}[NK: #1]}}
\else

  \newcommand{\NK}[1]{}
\fi

\begin{document}
\preprint{000-000-000}

\title{A flash before dawn:
21\,cm signatures of transient \\ reionization from supermassive Pop III stars}

\author{Nicholas Kern\,\orcidlink{0000-0002-8211-1892}}
\email{nkern@umich.edu}
\author{Prakhar Bansal\,\orcidlink{0009-0000-7309-4341}}
\author{Dragan Huterer\,\orcidlink{0000-0001-6558-0112}}%
\affiliation{
Department of Physics and Leinweber Institute for Theoretical Physics, University of Michigan, 450 Church St, Ann Arbor, MI 48109
}

\date{\today}

\begin{abstract}

Recent data from cosmological galaxy surveys combined with cosmic microwave background (CMB) data give a preference for a departure from $\Lambda$CDM, favoring a model with an evolving dark energy equation of state or a small neutrino mass. 
However, recent works have shown that the electron scattering optical depth to the surface of last scattering ($\tau$) can reduce this preference if it is allowed to take values of $\tau\sim0.09$, larger than that preferred by CMB low-$\ell$ E-mode polarization data.
With standard monotonic hydrogen reionization histories, such a high value of $\tau$ is in strong tension with spectroscopic galaxy data constraining the end of reionization.
Thus, high-$\tau$ models are restricted to non-standard reionization scenarios, where astrophysical sources inject an appreciable amount of ionizing photons at early times.
One recently studied model for this is a supermassive population III star scenario, where radiative feedback from a self-annihilating dark matter particle enables a pathway to form $M\sim10^5 M_\odot$ stars that, upon collapse, photoionize a surrounding volume large enough to source appreciable contributions to $\tau$ as a population.
Here we present the first self-consistent treatment of the feedback from these supermassive population III stars within a large semi-numerical cosmological simulation, accounting for their ionizing, Lyman-Werner, and X-ray feedback on the intergalactic medium. We use it to predict the distinct signatures on the large-scale 21\,cm signal at Cosmic Dawn, and forecast its detectability.
We find that the 21\,cm power spectrum probed by future radio survey arrays like the Square Kilometre Array could detect signatures of these supermassive population III at high significance and set model-dependent constraints on $\tau$ that approach $\sigma_\tau\sim0.012$, enough to distinguish our fiducial $\tau\sim0.097$ scenario from current Planck values at $>3\sigma$.

\end{abstract}

\maketitle



\section{Introduction}
\label{sec:intro}

Recently, new datasets from sensitive telescopes have challenged our assumptions regarding the universe's evolution and the formation of luminous objects at early times.
In particular, galaxy clustering data from the Dark Energy Spectroscopic Instrument (DESI) show preference for an evolving dark energy equation of state, reaching a $>2.5\sigma$ tension with standard $\Lambda$CDM when combined with external cosmological datasets \citep{DESIDR2}.
Additionally, the James Webb Space Telescope (JWST) has provided hints that early stellar populations may have produced more ionizing photons in the Cosmic Dawn era than originally thought and strengthened evidence for supermassive black holes at $z>7$ that are difficult to explain with conventional theories of black hole accretion \citep{Bogdan2024, Bogdan2024}.

One interesting model of high-redshift star formation may help to explain some of these disparate observations: supermassive population III star formation, modulated by dark matter annihilation radiative feedback, also known as dark stars.
Since its proposal \citep{Spolyar2008, Freese2008, Natarajan2009}, a number of authors have expanded upon the theory and debated its feasibility \citep[e.g.][]{Iocco2008, Smith2012, Stacy2014, Freese2016, Banik2019, Singh2023, Tan2024, Sanati2025, Petkova2026}.
These works show that, in the presence of annihilating dark matter, the first luminous objects forming in pristine gas in the cores of high-redshift dark matter halos could theoretically accrete to masses $\sim10^5\ M_\odot$.
Eventually, it is presumed these sources collapse onto the zero-age main sequence (ZAMS) and produce copious amounts of radiation over a short lifetime before directly collapsing into a massive black hole that could act as the heavy seeds needed to possibly explain the supermassive black hole population observed at $z>7$ \citep{Banik2019, Tan2024, Ilie2025}.
Recent works have shown that during their main sequence lifetime, these sources can photoionize a surrounding volume of $R_{\rm HII}\sim1.5$ cMpc \citep{Sanati2025, Tan2025flash}.
Thus, these sources could produce a \emph{transient hydrogen reionization} phase at high redshifts--distinct from standard galaxy-driven reionization at later times--where the intergalactic medium (IGM) is quickly partially ionized and then slowly recombines until standard galaxy-driven reionization takes over at $z < 9$.
The accretion and non-fusion supported phase of these objects has been referred to as a ``dark star'' based on the annihilation feedback that sustains these objects, while the short main-sequence lifetime after these objects collapse has also been referred to as supermassive Pop~III.1 stars \citep{Banik2019}.
This naming convention is intentionally overlapping with the more standard Pop~III.1 star definition: the first stars formed in ultra pristine environments, while Pop~III.2 stars formed in processed gas from prior Pop~III.1 stars \citep{McKee2008}.
Because this work is focused on the radiative output of these supermassive population III objects after they hit the ZAMS, we will refer to them as (supermassive) Pop~III.1 sources, with the implicit assumption that the underlying mechanism needed to reach supermassive states is driven by a form of radiative feedback from self-annihilation in the cores of dark matter halos.

In addition, considerable attention has been given recently to the robustness of the electron scattering optical depth parameter, $\tau$, tracing the integrated free electron fraction to the cosmic microwave background (CMB) surface of last scattering.
$\tau$ has long been known to act as a nuisance parameter that limits the ability of CMB data to precisely constrain $A_s$ \citep{Kaplinghat2003, Planck2020} and the neutino mass \citep{Allison2015, Sailer2026}.
Planck data has set very precise constraints on $\tau$ from the large-angle (low-$\ell$) E-mode polarization signal, which traces the hydrogen reionization history.
The Planck 2018 analysis prefers $\tau=0.054\pm0.007$ \citep{Planck2020}; however, omitting just the large-angle E mode data results in $\tau=0.08\pm0.012$ \citep{Giare2024}.
Furthermore, the robustness of the Planck low-$\ell$ E-mode polarization measurements has long been questioned given the complex polarized dust systematics from the galaxy and some known hardware systematics with Planck that impact the large-angle polarized maps, although to date there is no direct evidence for residual systematics in the Planck data \citep{PlanckInt2016XLVI, Delouis2019}.

If, however, we omit the low-$\ell$ E-mode polarization data, recent works have shown that a higher value of $\tau\sim0.09$ may alleviate some of the tensions with $\Lambda$CDM that DESI+CMB data have revealed \citep{Jhaveri2025, Sailer2026, Sullivan2026}, in particular the preference for dynamical dark energy \citep{DESIDR2} and a preference for a neutrino mass that is in tension with laboratory experiments \citep{Esteban2024}.
Thus, the value of $\tau$ and how to constrain it independently of the low-$\ell$ CMB polarization data has been an active area of recent inquiry.

Because $\tau$ is the mass-weighted integral of the free electron fraction along the line of sight, direct constraints on $\tau$ come from tracers of the electron field.
High-redshift quasar absorption spectra tracing the absorption of Lyman-$\alpha$ flux through the intergalactic medium (IGM) now fairly conclusively tell us that hydrogen reionization ended at $z\sim5.5$.
Thus, models that invoke a high $\tau$ must do so through early sources of ionization, either from excess population II output or outsized contributions from population III stars.
A few recent works have tried to pin $\tau$ without relying on Planck low-$\ell$ E mode data, using only Lyman-$\alpha$ probes of the midpoint and end of reionization, showing that $\tau$ is roughly constrained to be $<0.07$ \citep{Elbers2025, Kageura2026, MonteroCamacho2026}; however, these works generally assume a mostly neutral medium for $z > 12$, which may not be the case in alternative theories of structure formation \citep{Weiner2026}.

In particular, \citet{TanKomatsu2025, Aggarwal2026} have showed that the supermassive population III pathway may be able to boost the optical depth to $\tau\sim0.09$; however, these works did not implement supermassive Pop~III.1 feedback within a self-consistent cosmological simulation.
Modeling this feedback is important if we wish to make predictions for the large-scale 21\,cm signal emanating from the diffuse, neutral hydrogen medium in the IGM.
The 21\,cm signal has long been proposed as a powerful probe of cosmology and astrophysics at high redshifts \citep{Madau1997, Loeb2004, Furlanetto2006, Pritchard2012, Liu2020}.
In recent years, 21\,cm cosmology experiments have dramatically matured and are delivering novel scientific insights into the astrophysical and cosmological physics of the high redshift universe \citep{HERA2022a, HERA2023, Nunhokee2025, Mertens2025, CHIME2026, MeerKat2026}.
Going forward, next-generation 21\,cm experiments like the Square Kilometre Array \citep{Braun2019} will push on the redshift and sensitivity frontier for 21\,cm cosmology, making it necessary to develop new predictive models of high-redshift astrophysical structure formation.

In this work, we present the first self-consistent implementation of supermassive population III feedback on the large-scale IGM within a semi-numerical cosmological simulation, specifically the \texttt{21cmFAST} package \citep{Mesinger2011, Park2019, Qin2020, Murray2020, Davies2025}.
We model the ionizing, Lyman-Werner, and X-ray output of these sources over time, accounting for the inhomogenous ionization, recombination, heating, and cooling of the IGM.
We confirm that this model of supermassive population III stars can result in a high value of $\tau$ that is still consistent with low-redshift Lyman-$\alpha$ data constraining the end of galaxy-driven hydrogen reionization.
We use this model to track the inhomogenous 21\,cm spin temperature and make predictions for 21\,cm observables, including the power spectrum and global signal, and produce forecasts for detectability with the Square Kilometre Array Low (SKA-Low).
The goal of this work is not to determine the feasibility of dark star or supermassive population III star formation.
Instead, we are interested in determining their signatures on current and future observables if these objects exist as prescribed in the literature.
More broadly, we are interested in asking the question: if a population of ionized bubbles at high redshift ($z>15$) are sourced by a mechanism that traces star formation, can 21\,cm observables detect them and infer $\tau$?

In a companion paper, we also study the impact that these supermassive Pop~III.1 sources have on the kinetic Sunyaev-Zeldovich (kSZ) signal, showing it to be another probe of high-redshift contributions to the optical depth at $\tau(z>12)$ \citep{Bansal2026}.
In \S2 we describe the PopIII.1 feedback model and our semi-numerical implementation in more detail.
In \S3 we present a forecast of the signal's potential detectability with the SKA-Low, and in \S4 we discuss the implications of this work, its limitations, and future extensions.

\section{Supermassive population III model}
\label{sec:pop31}


We implement supermassive PopIII.1 source placement and radiative feedback within the semi-numerical cosmological simulation \texttt{21cmFAST} v4 \citep{Mesinger2011, Greig2015, Park2019, Qin2020, Murray2024, Davies2025}.
We use the standard excursion set theory in 21cmFAST for galaxy-driven reionization with inhomogenous recombinations, mini-halo ionization and heating, and track X-ray, Ly-$\alpha$ and CMB heating of the IGM (all standard 21cmFAST features).
We simulate 3D boxes with sidelength $L=252$ cMc on a 504$^3$ grid (with a 1008$^3$ grid for the initial conditions), a voxel resolution of 0.5 cMpc, and evolve the fields down from $40 > z > 5.5$ with log-spaced reshifts using a step factor of 1.02.
We ran a number of numerical tests to ensure our fiducial set of simulation runs were converged given the stated approximate physics of our semi-numerical model.
The two major convergence tests we ran were ensuring the 21\,cm observables were stable given the adopted cell resolution and the redshift time stepping resolution.
For the former, we ran three simulations with cell sizes of 1.0, 0.75, and 0.5 cMpc, finding $<5\%$ convergence in our observables-of-interest at the smallest resolution, which is what we adopt for our production runs.
We found it important to resolve the $R_{\rm HII}$ scale by at least a factor of 2, which our 0.5 cMpc resolution exceeds with our fiducial set of Pop~III.1 parameters yielding $R_{\rm HII}\sim1.5$ cMpc.
Similarly, we found $<5\%$ convergence in the various observables with a log redshift step factor of 1.02 (which is 21cmFAST's default value), relative to factors of 1.03 and 1.04.

For galaxy parameters, our starting point is the \citet{Qin2020} 21cmFAST template with Planck \citep{Planck2020} cosmological parameters.
What differs in our base model from that of \citet{Qin2020} are: 1) we account for relative velocities between baryons and dark matter in star forming halos, 2) we account for CMB and Ly-$\alpha$ heating of the IGM, 3) we slightly decrease the minihalo radiative output to calibrate the galaxy-only optical depth to $\tau\sim0.06$, and we only allow for minihalo formation within previously flash ionized regions (described below), to be consistent with the PopIII.1 -> PopIII.2 pathway (i.e. we use 21cmFAST's minihalo model as an effective PopIII.2 population).
The total runtime for a single sweep from $40>z>5.5$ with our new PopIII.1 physics included is increased by only a factor of 1.6x on a CPU node.
We describe the major additions to the 21cmFAST codebase below. The code will be made publicly available at a future date with a publication detailing the software.

\subsection{PopIII.1 Source Formation}
\label{sec:placement}

A supermassive PopIII.1 star is assumed to be the first object formed in its vicinity in the core of a high-redshift dark matter halo.
If the dark matter is a self-annihilating weakly interacting massive particle (WIMP), the resulting gas heating rate from its annihilation byproducts can halt standard Kelvin-Helmholtz contraction \citep{Spolyar2008, Freese2008}.
A conventional Pop~III protostar contracts to the main sequence and its ionizing feedback halts accretion at $M\sim10^2\ M_\odot$ \citep{McKee2008, Bromm2009}.
A ``dark star,'' by contrast, stays large and cool at $T_{\rm eff}\le10^4$ K, is not supported by hydrogen fusion and does not emit substantial ionizing radiation, and can grow to up to $M\sim10^5M_\odot$ over an accretion time $t_{\rm acc}\sim10-100$ Myr \citep{Freese2016}.

After collapse onto the ZAMS and fusion ignition, the Pop~III.1 star lives for $t_{\rm life}\sim8$ Myr, during which it emits a hard ionizing spectrum at a rate of $Q_{\rm ion}\sim3\times10^{53}\ {\rm s}^{-1}$ photons.
Over the duration of the star's life, this carves out a photoionized sphere around the source of radius $R_{\rm HII}$.
Because the Pop~III.1 models assumes these sources can only form in pristine, undisturbed gas, once a region has been flash ionized by a Pop~III.1 source it can no longer form one in the future.
Historically, this was referred to as an isolation distance, $d_{\rm iso}$, which can be tuned to set the overall Pop~III.1 remnant density \citep{Banik2019}.
However, physically speaking this should be tied to the radius of the HII region, with perhaps an additive reach to account for effects such as momentum-driven bubble expansion and partial ionization.
Even small amounts of partial ionization can create cooling pathways that would suppress supermassive Pop~III.1 formation \citep{Johnson2006}.
To account for these unmodeled subgrid effects, we define $d_{\rm iso} = a_{\rm iso}R_{\rm HII}$, with a default value of $a_{\rm iso}=1.4$ calibrated to numerical simulations \citep{Sanati2025}.
This yields a Pop~III.1 remnant density of $n_{\rm III.1}\sim 0.075\ {\rm cMpc}^{-3}$, which is roughly consistent with previous works.

\begin{figure}[!t]
\centering
\includegraphics[width=\linewidth]{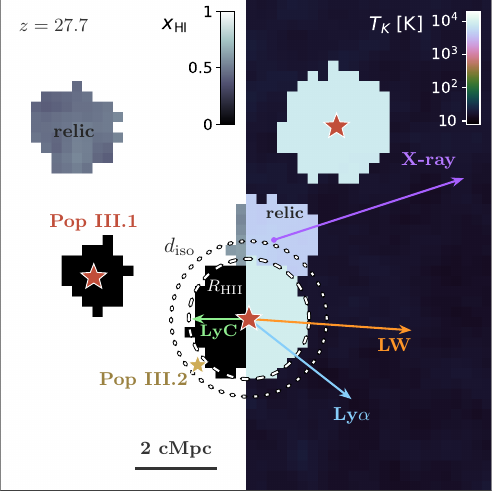}
\caption{
Isolation-regulated placement and the radiative channels of Pop~III.1 sources, simulated at 0.25 cMpc cell resolution.
We plot the hydrogen neutral fraction field (left), showing ionized bubbles with a radius $R_{\rm HII}$ around actively emitting Pop~III.1 sources (red star), alongside recombining relic bubbles after the sources turn off.
During their emission phase, Pop~III.1 sources emit ionizing photons (LyC), Lyman-$\alpha$ photons that scatter throughout the neutral IGM, and Lyman-Werner photons that dissociate H$_2$, making neighboring Pop~III.1 and Pop~III.2 formation less efficient.
At later times, it is assumed the remnant black holes go through accretion phases and emit X-rays that penetrate the IGM with long mean-free paths, dumping energy and partially ionizing the simulation volume over time.
Each source creates an exclusion zone of radius $d_{\rm iso}$ around it, preventing future sources to form; however, only within $d_{\rm iso}$ are Pop~III.2 sources allowed to form in the future.
Bubbles with active sources have a kinetic gas temperature $T_{\rm relic}$ (right), which then begins to cool after the sources turn off.
}
\label{fig:diso}
\end{figure}

\begin{figure*}[!t]
\centering
\includegraphics[width=\linewidth]{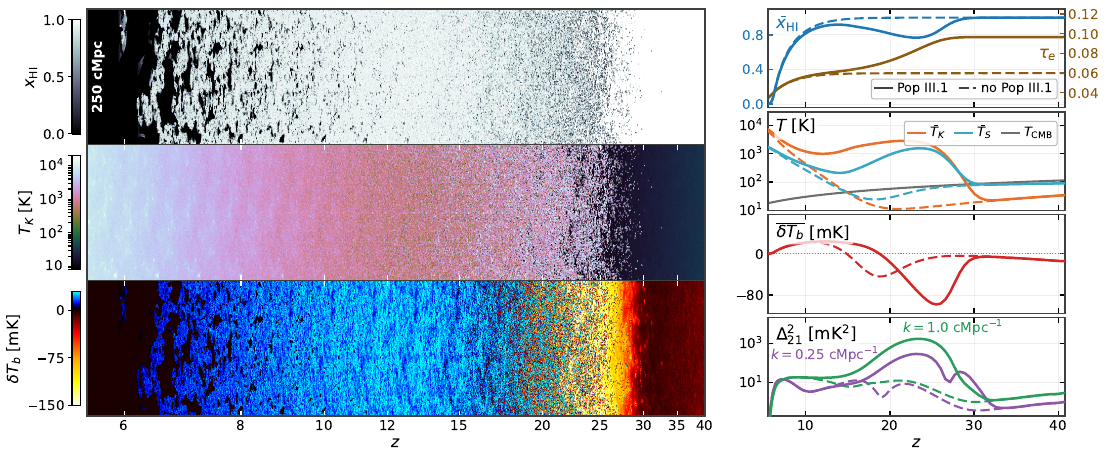}
\caption{
Lightcones from our fiducial Pop~III.1 run, showing the hydrogen neutral fraction field ($x_{\rm HI}$; top), the gas kinetic temperature ($T_K$; middle), and the 21\,cm brightness temperature ($\delta T_b$; bottom). We also plot their globally-averaged profiles (right), the optical depth $\tau(z)$, and the 21\,cm power spectrum $\Delta^2_{21}$ with and without (dashed) Pop~III.1 physics.
}
\label{fig:summary}
\end{figure*}

Minihalos are not resolved by the grid, so the rate at which they collapse is taken from the extended Press--Schechter (EPS) conditional collapse fraction ($f_{\rm PS}$) \citep{PressSchechter1974,Bond1991,LaceyCole1993}, evaluated with \texttt{21cmFAST}'s density field RMS output $\sigma(M)$ and growth factor $D(z)$ (no discrete halo catalogue is used).
As in 21cmFAST, we re-scale the collapse fraction cell-by-cell to account for rare halos,
\begin{equation}
f_{\rm coll}=r(\nu)\,f_{\rm PS},\qquad r(\nu)=\frac{F_{\rm W}(>\nu,z)}{\mathrm{erfc}(\nu/\sqrt2)},
\label{eq:stnorm}
\end{equation}
where $\nu=\delta_c/[D(z)\sigma(M_{\rm eff})]$, $\delta_c=1.686$, and $F_{\rm W}$ is the Watson mass fraction above $\nu$, with \texttt{21cmFAST}'s redshift-dependent coefficients \citep{Watson2013}.
We use the Watson fit rather than Sheth--Tormen (ST) because it is calibrated on N-body simulations at these redshifts and agrees to $30\%$ with the minihalo mass function \citep{Reed2007}.
Galaxies and Pop~III.2 minihalos keep \texttt{21cmFAST}'s ST normalization.

The minimum mass of a halo hosting a Pop~III.1 star must be at least $M_{\rm min}\sim10^6 M_\odot$ to build a supermassive star \citep{Banik2019}, and is further modulated by baryon--dark matter streaming velocity ($v_{cb}$) and H$_2$ dissociating Lyman-werner feedback \citep{Haiman2000, Machacek2001, Visbal2014, Fialkov2014, Munoz2022}.
We model this the same way 21cmFAST does as a minimum effective host mass 
\begin{align}
\label{eq:Meff}
M_{\rm eff} = {\rm max}[M_{\rm min}, m_{\rm crit}(z)f_{\rm LW}f_{\rm vcb}],
\end{align}
where $f_{\rm LW}=1 + A_{\rm LW}J_{\rm 21,LW}^{\beta_{\rm LW}}$, $f_{\rm vcb}=(1+A_{\rm vcb}\tfrac{v_{cb}}{\sigma_{\rm vcb}})^{\beta_{\rm vcb}}$, and $m_{\rm crit}\propto(1+z)^{-1.5}$ is the H$_2$ cooling mass in 21cmFAST.
$A_{\rm LW}=0.35$, $\beta_{\rm LW}=0.6$, $A_{\rm vcb}=1.0$, $\beta_{\rm vcb}=1.8$, and $\sigma_{\rm vcb}=29$ km s$^{-1}$ are LW and streaming velocity parameters inherited from 21cmFAST.
$J_{21,\rm LW}$ is the Lyman-Werner background incident on a cell, computed in 21cmFAST by summing along the past lightcone.

At every reshift and each pristine (non-excluded) grid cell, we draw the number of Pop~III.1 sources formed at that redshift and grid cell as $N\sim{\rm Poisson}(\lambda_n)$, where 
\begin{align}
\label{eq:lambda}
\lambda_n(\boldsymbol{x},z) = \frac{M_{\rm cell}\Delta f_{\rm coll}(\boldsymbol{x},z)}{M_{\rm eff}}  
\end{align}
is the number of host halos above the minimum effective mass that collapsed during the intervening time $\Delta z$, and $\Delta f_{\rm coll}$ is the difference in the collapse fraction from $z_{i-1}$ to $z_{i}$.
For each of the $N\sim{\rm Poisson}(\lambda_n)$ Pop~III.1 sources drawn for a given grid cell we assign a random position within the grid cell.
We then compute its distance from all previously-formed Pop~III.1 sources and if it falls outside of their $d_{\rm iso}$ we consider it formed and add its coordinates to the catalogue.
We draw all cell's $N$ simultaneously, randomly order them and proceed with this process iteratively, meaning that a new Pop~III.1 source can suppress its neighboring cell's newly formed sources within the same redshift step.
This way of drawing sources make the physics cell-resolution-independent on average, which is desireable because we run the simulation at a fairly coarse resolution that resolves the ionized bubbles by a factor of $\sim3$ (convergence is verified).
Individual cell's are also excluded if they have ionization fractions $x_e\ge10^{-3}$, which is eventually achieved throughout the box from X-ray ionizations, acting as a natural shutoff at late times \citep{Johnson2006, Yoshida2007, Hummel2015}.

To account for the accretion time between a halo collapsing and its ZAMS Pop~III.1 window--what we call $t_{\rm acc}$--we actually evaluate \autoref{eq:lambda} at $z + \Delta z(t_{\rm acc})$, which delays Pop~III.1 physics by the appropriate amount of time.
We assume no ionizing, Lyman-$\alpha$, Lyman-Werner or X-ray emission during this time.
\autoref{fig:diso} shows a schematic of Pop~III.1 source formation (red star), showing the response of the surrounding neutral fraction and temperature field.
During the active phase, source emit ionizing, Lyman-$\alpha$, and Lyman-Werner radiation.
At some later point after sources turn off, the remnant black holes accretes gas and begins to emit X-rays that penetrate the IGM, heating and partially ionizing it, which eventually shuts off all future Pop~III.1 formation within the simulation volume.

The impact of the isolation procedure is that, at early times, Pop~III.1 sources form in highly biased regions (the rare peak overdensities), but these regions eventually become isolated.
Thus overtime Pop~III.1 source placement becomes a packing problem, and sources form in progressively less-biased regions that take longer to cross the minimum hosting mass \citep{Bansal2026}.
In our fiducial runs, Pop~III.1 sources begin forming at $z\sim33$, quickly reach peak $dn/dz$ formation rates at $z=24$, and then slowly decrease over time as they isolate most overdense regions with 90\% formed by $z\sim19$, until formation is completely shutoff at $z\sim15$ through X-ray feedback. 

\subsection{Ionization and thermal evolution of IGM}
\label{sec:feedback}

Once a Pop~III.1 source forms via the mechanism described above, all cells within its $R_{\rm HII}$ are completely ionized.
This is modeled as an R-type ionization front, with a radius that is set by ionizing photon counting against the mean HI density, which in co-moving coordinates cancels out and we are left with
\begin{align}
R_{\rm HII} = 1.47\ {\rm cMpc}\ \left(\frac{Q_{\rm ion}}{3\times10^{53}\ {\rm sec}^{-1}}\right)^{1/3}\left(\frac{t_{\rm life}}{8\ {\rm Myr}}\right)^{1/3},
\end{align}
independent of redshift and with a weak dependence on local oversensity $(1+\delta)^{-1/3}$ that we omit \citep{Tan2025flash}.
For normal massive Pop~III star emitting near its Eddington luminosity, the nuclear burning lifetime is $\sim2$ Myr \citep{Woods2020}, and is roughly masss independnet.
Longer lifetimes of $2-20$ Myr are adopted within the Pop~III.1 literature \citep{Tan2025flash} to account for residual WIMP annihilation heating that can prolong the stellar lifetime \citep{Spolyar2008, Natarajan2009}.

While a source is active, all cells within $R_{\rm HII}$ have $x_e=1.0$, and photoheating balanced by CMB Compton cooling sets the kinetic temperature of the bubble at $T_{\rm eq}=T_{\rm relic}(1+25)/(1+z)$ K, where $T_{\rm relic}=10^4$ K is the relic temperature at $z=25$ \citep{Tan2025flash}.
Cells that bisect $R_{\rm HII}$ are randomly chosen to be fully ionized or fully neutral proportional to the volume overlap (21cmFAST's ionization code assumes a single-phase media).
Convergence tests confirm this strategy keeps the 21\,cm observables independent of choice of cell resolution, although it is required that the grid scale be at least $2\times$ smaller than the ionized bubble size $R_{\rm HII}$.

Ionized gas contributes directly to the electron scattering optical depth by scattering free-streaming CMB photons.
It is an integral quantity proportional to the mass-weighted electron density, written as
\begin{equation}
\tau(z^\prime)=c\,\sigma_T\,n_{\rm H,0}\int_0^{z^\prime}\Gamma_{\rm He}(z)\,\overline{(1+\delta)\,x_{\rm HII}}(z)\,
\frac{(1+z)^2}{H(z)}\,dz ,
\label{eq:tau}
\end{equation}
where $\overline{(1+\delta)\,x_{\rm HII}}$ is the mass-weighted hydrogen ionized fraction, $n_{\rm H,0}$ is the present-day mean hydrogen density, and $\Gamma_{\rm He}(z)=[1+f_{\rm He}+f_{\rm He}x_{\rm HeIII}(z)]$ is the additional free electrons from single and doubly ionized Helium \citep{Liu2016}.
As in most works, we assume HeII occurs simultaneously with HII, while $x_{\rm HeIII}(z)$ is modeled as a tanh function at $z=3$ with $\Delta z=0.5$, and $f_{\rm He}=0.079$ is the Helium fraction.

Once the gas is flash ionized and the Pop~III.1 source turns off the bubble becomes a relic and begins to cool and recombine as a coupled system,
\begin{align}
\label{eq:cooling}
\dot{x}_e &= -\mathcal{R}=-C_{\rm HII}\,\alpha_A(T_K)\,n_{\rm H}(z)\,(1+\delta)x_e^2 \\
\begin{split}
\dot{T}_K &= -2H(z)\,T_K - \frac{x_e(T_K-T_\gamma)}{(1+f_{\rm He} + x_e)t_C} \\
&\,\quad+ \frac{(1-\tfrac{2}{3}\beta(T_K))T_K}{1+x_e}\mathcal{R},
\end{split}
\end{align}
where $C_{\rm HII}=4$ is the sub-cell clumping factor, $\alpha_A(T_K)$ is the case A recombination coefficient, $T_\gamma$ is the background CMB temperature, $t_C$ is the Compton cooling time, and $\beta(T_K)$ is the thermal energy radiated per recombination \citep{HuiGnedin1997}.
The clumping factor is not well constrained, but it's fiducial value is consistent with expectations for flash-ionized, high-redshift gas \citep{DAloisio2020, Tan2025flash}.
The terms in the heating equation are adiabatic cooling, inverse Compton cooling off the CMB, and net recombination heating (the $n_e$ recombination drop outweighs radiative recombination cooling), respectively.
For consistency with 21cmFAST, we only consider case A recombinations (optically thin).
Differences between case A and case B can be effectively be folded into the clumping factor, which we keep as a free parameter.

In 21cmFAST, this set of physics is tracked and evolved in the X-ray module, which does account for mixed-phase media: ionized and neutral gas co-exist in an actively recombining Pop~III.1 relic bubble, which 21cmFAST's ionization code cannot handle.
However, X-ray ionizations rarely reach above $x_e>10^{-2}$, so 21cmFAST uses a simplified Euler solver for \autoref{eq:cooling}, which in our case is insufficient because of large changes in $x_e$.
Furthermore, they don't account for recombination cooling as it's largely negligible for $x_e<10^{-2}$, however, we include it in \autoref{eq:cooling} for our larger ionized fractions.
To solve \autoref{eq:cooling} we instead split each redshift step into 16 sub-steps (for each grid cell), holding the coefficients constant, and employ a solver with second-order accuracy. Convergence tests confirm the accuracy of this approach, without which the relic gas would cool 7x too fast and recombine 5x too slow.

\begin{figure*}[!t]
\centering
\includegraphics[width=\textwidth]{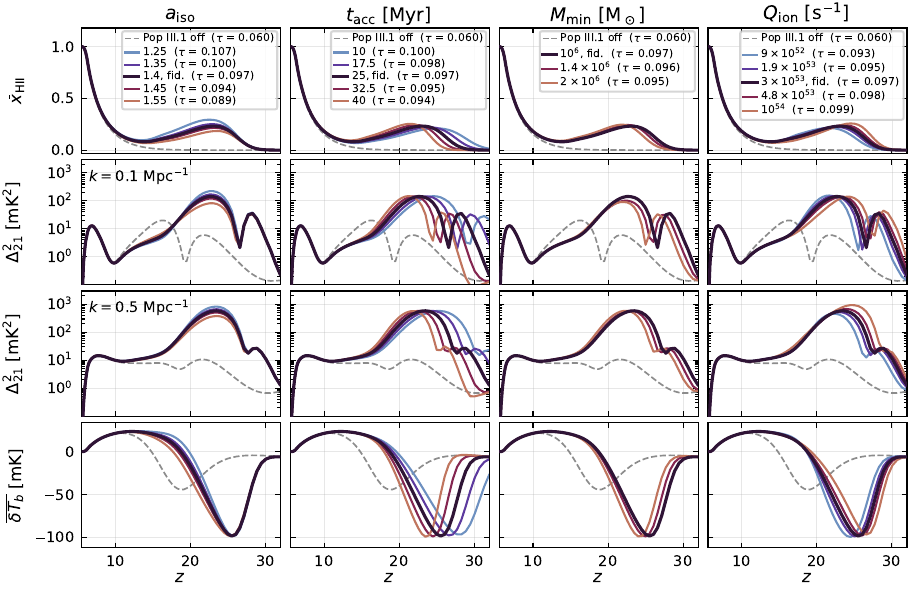}
\caption{
Response of the model to four of the main Pop~III.1 semi-numerical parameters determining 1) the packing density, modulating the height of the ionized fraction, 2) Pop~III.1 ignition delay, modulating the timing, 3) minimum source host mass, also modulating timing, 4) ionizing photon output, modulating timing and ionized fraction height.
Pushing the ionized fraction higher, wider, and to earlier redshifts increases $\tau$.
Parameters that can sharpen the ionization bump in $x_{\rm HII}$ include reducing Lyman-Werner feeback (which draws out when sources form in time), and increasing the clumping factor (which accelerates recombinations).
For $z<10$ all observables shown are similar to their Pop~III.1-off counterpart.
}
\label{fig:sweep}
\end{figure*}

\subsection{Lyman-band radiation}
\label{sec:lyman}

Pop~III.1 sources also emit photons between Lyman-$\alpha$ and the Lyman limit (10.2 -- 13.6 eV).
These photons drive Wouthuysen-Field (WF) coupling \citep{Wouthuysen1952,Field1958,Hirata2006} with Lyman-series cascades, drive Ly-$\alpha$ heating \citep{Chen2004, Chuzhoy2007, Reis2021}, and dissociate H$_2$, raising the molecular cooling turnover \citep{Haiman2000, Visbal2014}.
The spin temperature of the 21\,cm transition sets the relative occupancy of the singlet and triplet state: $n_1/n_0 = \exp\left[-T_\star/T_S\right]g_1/g_0$, where $T_\star=h\nu_{21}/k_B$=0.068 K is the 21\,cm transition energy.
The spin temperature is coupled to the radiative, collisional, and Lyman-$\alpha$ reservoirs,
\begin{align}
T_S^{-1} = \frac{T_\gamma^{-1} + x_cT_K^{-1} + x_\alpha T_\alpha^{-1}}{1 + x_c + x_\alpha},
\end{align}
where $x_c$ and $x_\alpha$ are collisional and Lyman-$\alpha$ coupling coefficients, $T_K$ is the kinetic temperature of the gas, and $T_\alpha$ is the color temperature.

The collisional coupling mechanisms are through H-H, $e^-$-H, and $p^+$-H collision rates and the local baryon density, with a redshift scaling of $x_c\propto(1+z)^2$ \citep{Mesinger2011}.
The WF coupling coefficient is set by the local angle-averaged Lyman-$\alpha$ intensity, written as
\begin{align}
x_\alpha=\frac{16\pi^2 e^2f_\alpha}{27A_{10}m_e c}\frac{T_\star}{T_\gamma}S_\alpha J_\alpha,
\end{align}
where $A_{10}=2.85\times10^{-15}$ s$^{-1}$, and $f_\alpha$ is the Ly-$\alpha$ oscillator strength \citep{Qin2020}.
$S_\alpha$ is a suppression factor fit accounting for repeated scattering \citep{Chen2004, Hirata2006}.
$J_\alpha$ is the angle-averaged Ly-$\alpha$ specific intensity as is defined in 21cmFAST, with contributions from galaxies, minihalos, and in this work from our Pop~III.1 sources.
We model the Lyman-band output as $Q_{\rm Ly}=f_{\rm Ly}Q_{\rm ion}$ per source, with a power-law spectrum $dN_\gamma/d\nu\propto\nu^{\alpha_{\rm Ly}}$ normalized over the band. Our fiducial model takes a hot-ZAMS value, $f_{\rm Ly}=0.3$ and $\alpha_{\rm Ly}=1$.
The same spectrum contributes to the Lyman-Werner background, which raises the minimum halo mass for star formation (\autoref{eq:Meff}).
Note that photon propagation is not explicitly computed: the off-resonant continuum is assumed optically thin in the IGM until it reshifts into the resonant frequencies.
This is the standard treatment in 21cmFAST \citep{Qin2020}, we simply add new Pop~III.1 sources in the angle-averaged Lyman-band emissivities.
Unlike the standard treament, however, we turn off these sources' Lyman-band emission after their lifetime.

We also considered recombining relic HII regions as an additional source of Lyman-$\alpha$ photons.
These Ly-$\alpha$ photons are created at line center from recombinations and therefore only interact with the gas in the emitted grid cell, similar to 21cmFAST's X-ray-excited Lyman-$\alpha$ treatment, and are emitted over the lifetime of the relic bubble.
However, because the relic gas is largely already WF-coupled, this effect turns out to be small.

\subsection{Photoheating feedback in relic HII regions}
\label{sec:photoheating}

Heated gas cannot accrete into a halo until it falls below its virial temperature.
Relic HII regions therefore cannot host future Pop~III.2 and Pop~II stars until it has recombined and cooled to some degree.
21cmFAST reionization photoheating feedback is folded into the excursion set ionization module.
In it, the code exponentially supresses star formation in halos with $M<M_{\rm turn}$ \citep{Qin2020}, where $M_{\rm turn}=M_{\rm crit}^{\rm reion}$ is computed from numerical simulations of reionization \citep{Sobacchi2014}.
However, as noted we use it's alternative X-ray module to handle relic heating and cooling.
We thus compute a photoheating suppressed halo mass for each relic grid cell based on its temperature and ionization state, $M_{\rm crit}^{\rm relic}=M_{\rm vir}(z,T_K, x_e)$, and fold it into the excursion set photoheating prescription--$M_{\rm turn}={\rm max}[M_{\rm crit}^{\rm reion},M_{\rm crit}^{\rm relic}]$--before computing Pop~III.2 and Pop~II radiative properties.

\begin{table*}[!t]
\caption{Parameters varied in the Fisher forecast, their fiducial values and their Gaussian priors ($1\sigma$). All fiducial values quoted in linear units.}
\label{tab:fisher_params}
\begin{ruledtabular}
\begin{tabular}{llll}
Parameter & Description & Fiducial (linear) & Prior ($1\sigma$) \\
\hline
\multicolumn{4}{l}{\emph{Pop~III.1 source parameters}}\\
$a_{\rm iso}$ & isolation distance coefficient & 1.4 & 0.5~\cite{Sanati2025, Petkova2026} \\
$\log_{10}t_{\rm acc}$ & dark accretion phase before ZAMS & 25\,Myr & 0.6 dex~\cite{Sanati2025} \\
$\log_{10}A_{\rm LW}^{\rm III.1}$ & Lyman--Werner feedback on $M_{\rm crit}$ & 0.35 & 0.75 dex \\
$\log_{10}t_{\rm life}$ & luminous lifetime & 8\,Myr & 0.4\,dex~\cite{Sanati2025, Petkova2026} \\
$\log_{10}M_{\rm min}$ & minimum halo host mass & $10^6\,M_\odot$ & 1.0\,dex~\cite{Sanati2025, Petkova2026} \\
$\log_{10}Q_{\rm ion}$ & ionizing photon rate & $3\times10^{53}\,{\rm s^{-1}}$ & 0.75\,dex~\cite{Sanati2025, Petkova2026} \\
\hline
\multicolumn{4}{l}{\emph{Pop~III.1 nuisance parameters}}\\
$\log_{10}C_{\rm HII}$ & sub-cell clumping within relic bubbles & 4 & 0.25\,dex~\cite{DAloisio2020,TanKomatsu2025} \\
$\log_{10}T_{\rm relic}$ & initial relic temperature at $z=25$ & $10^4$\,K & 0.3\,dex~\cite{Sanati2025, Petkova2026} \\
$\log_{10}f_{\rm Ly}$ & Pop~III.1 Lyman-band photons per ionizing photon & 0.3 & 0.3\,dex \\
$\log_{10}f_{X}$ & Pop~III.1 remnant soft-band X-ray output  & 0.03 & 0.5\,dex~\cite{Duras2020} \\
\hline
\multicolumn{4}{l}{\emph{Galaxy and minihalo parameters}}\\
$\log_{10}f_{*,10}$ & stellar fraction at $10^{10}\,M_\odot$ & 0.056 & 0.3\,dex~\cite{Park2019} \\
$\alpha_*$ & stellar fraction mass slope & 0.6 & 0.07~\cite{Park2019} \\
$\log_{10}f_{\rm esc,10}$ & galaxy ionizing escape fraction & 0.048 & 0.75\,dex \\
$\log_{10}f_{*,7}^{\rm mini}$ & minihalo stellar fraction & 0.018 & 0.75\,dex \\
$\log_{10}A_{\rm LW}$ & minihalo Lyman--Werner feedback & 2.0 & 0.5 dex~\cite{Schauer2021,Munoz2022} \\
$\log_{10}L_X$ & galaxy X-ray luminosity per SFR & $3.2\times10^{40}$ erg\,s$^{-1}$ & 1.0\,dex~\cite{Qin2021} \\
$\log_{10}L_X^{\rm mini}$ & the same for minihalos & $10^{41}$ erg\,s$^{-1}$ & 1.0\,dex~\cite{Qin2021} \\
$E_0$ & X-ray self-absorption cutoff & 500\,eV & 400\,eV~\cite{Das2017} \\
\end{tabular}
\end{ruledtabular}
\end{table*}

\subsection{Proto-AGN X-ray heating and ionization}
\label{sec:x_rays}

After the Pop~III.1 source runs its course on the main sequence, it directly collapses into a black hole.
After collapse, these sources lie dormant for some period until blown-out material begins to accrete onto the compact object, forming an accretion disk that radiates in the X-ray.
The soft band component can penetrate and heat the IGM over its long mean free path.
This could be important in predicting the 21\,cm signal that is sensitive to the gas temperature.
In our model, each Pop~III.1 source leaves a black hole with $M\sim10^5M_\odot$ after it's ZAMS lifetime, after which it lies dormant for $t_{\rm delay}\sim 20$ Myr until the expelled gas begins to accrete and emit X-rays \citep{Sanati2025, Sanati2025b}.
We model the time-averaged bolometric luminosity as
\begin{align}
L_X = f_{\rm X}f_{\rm duty}\lambda_{\rm Edd} L_{\rm Edd}(M),
\end{align}
where $f_X=0.03$ is the fraction emitted in the soft band, $f_{\rm duty}=5\times10^{-3}$ is the fraction of time the source spends in an accretion phase, $\lambda_{\rm Edd}=L_{\rm bol}/L_{\rm edd}=0.3$ is the Eddington ratio, and $L_{\rm Edd}=1.26\times10^{38}(M/M_\odot)$ erg s$^{-1}$ \citep{Duras2020,Cammelli2025, Sanati2025b}.
We model the soft band spectral luminosity as a power-law, $L\propto E^{-\alpha_X}$, with a sharp cutoff at low energies, normalized to $L_X$,
\begin{align}
L_X = \int_{E_0}^{\rm 2\, keV}\, L\, dE,
\end{align}
where $E_0=500$ eV and $\alpha_X=1$ \citep{Yang2015, Sanati2025}.
This is fed then into the standard 21cmFAST X-ray code, which heats and (slightly) ionizes the cold IGM \citep{FurlanettoStoever2010, Mesinger2013, Qin2020}.
In our fiducial runs, we find that X-ray heating of the cold IGM by Pop~III.1 remnants is most important for $14 < z < 22$: after these proto-AGN have turned on but before galaxy-driven X-ray heating takes over.
In this range, Pop~III.1 remant X-ray heating changes the mean gas temperature (of non-relic regions) by $\sim40\%$, reduces the trough of the global signal by $\sim5\%$, and lowers the 21\,cm power spectrum by $\sim20\%$.
Uncertainty in these X-ray SED prescriptions is large, but to first order we can encapsulate the impact of Pop~III.1 X-ray heating by simply varying the overall soft band luminosity, $f_X$.

\subsection{The 21\,cm signal}
\label{sec:21cm}

The differential 21\,cm brightness temperature in the optically thin limit is
\begin{align}
\label{eq:21cm}
\delta T_b(z) \simeq \delta T_0\,x_{\rm HI}\,(1+\delta)\left(1-\frac{T_\gamma}{T_S}\right)\ {\rm mK},
\end{align}
where $\delta$ is the matter overdensity, $x_{\rm HI}$ is the neutral fraction, $\delta T_0=27\left(\Omega_bh^2/0.023\right)\left[(0.15(1+z))/(10\Omega_mh^2)\right]^{1/2}$, and we've ommited the line-of-sight velocity gradient factor for simplicity \citep{Furlanetto2006, Pritchard2012, Liu2020}.
The spatial average, $\overline{dT_b}(z)$, is known as the global signal (i.e. the monopole), and carries information on the timing of source formation, WF coupling, X-ray heating, and reionization \citep{Shaver1999, Pritchard2010, Pritchard2012}.
The spatial fluctuations of the signal can be encapuslated in its power spectrum,
\begin{align}
(2\pi)^3\delta_D({\bf k}-{\bf k}^\prime)P_{21} = \langle \tilde{T}({\bf k})\tilde{T}({\bf k}^\prime)^\ast\rangle,
\end{align}
where $\tilde{T}({\bf k})$ is the 3D Fourier transform of the field with wavevector ${\bf k}=(k_x, k_y, k_z)$, and $\delta_D$ is the Dirac delta function.
We define the dimensionless power spectrum as
\begin{align}
\Delta^2(k, z) = \frac{k^3}{2\pi^2}P_{21}(k,z)\ {\rm mK}^2.
\end{align}
The power spectrum carries information regarding the clustering of the density, ionization, and temperature of the IGM, bearing signatures of Pop~III.1 sources and their feedback.
Due to the non-linear processes of source formation and ionization, the 21\,cm field in theory carries information beyond its 2-point function; however, in this work we only present forecasts for the global signal and power spectrum.

\autoref{fig:sweep} shows the response of these observables, along with the ionization fraction $x_{\rm HII}=x_e$, to perturbations of key Pop~III.1 parameters, including the isolation criterion, $a_{\rm iso}$, the accretion lifetime $t_{\rm acc}$, the minimum halo mass $M_{\rm min}$, and the ionization rate, $Q_{\rm ion}$.
Roughly speaking, these control the timing of the transient reionization phase and the amount of the IGM that gets ionized.
This manifests as modulating the 21\,cm observables forward and backward in redshift, and a slight increase or decrease in overall power.
The power spectrum shows two major bumps as a function of redshift that correspond to early Ly-$\alpha$ coupling ($z>25$) and the contrast between hot relic bubbles and the cold IGM ($z < 25$), which increases the power spectrum amplitude at $z\sim23$ by nearly two orders of magnitude.
One of the unique signatures introduced by Pop~III.1 stars (and probably any major star forming event at very high redshifts) is the early onset of Lyman-$\alpha$ coupling of the 21\,cm spin temperature.
This drives the global signal to absorption much sooner than in the standard Pop~III case (dashed grey), which is difficult to re-produce even when accounting for uncertainty in Pop~II and Pop~III formation \citep{Cohen2020}.
By $z < 10$ where we have direct probes of the IGM, the neutral fraction and 21\,cm observables are consistent with standard physics.

\section{Forecasting parameter constraints}
\label{sec:forecast}

Here we investigate whether an upcoming 21\,cm experiments could detect the fiducial signal predicted by our simulations (\autoref{fig:sweep}).
We produce forecasts for the Square Kilometre Array Low (SKA-Low) \citep{Koopmans2015} and a global signal experiment, such as the Radio Experiment for the Analysis of Cosmic Hydrogen (REACH) \citep{deLeraAcedo2022}.
We use a Fisher matrix formalism to forecast the projected uncertainty on astrophysical parameters and derived quantities like $\tau$ and $x_{\rm HII}$.
In \autoref{tab:fisher_params} we tabulate all of the free parameters in our forecast, including the most relevant Pop~III.1 source placement and HII bubble parameters, secondary parameters that do not directly impact Pop~III.1 sources but do influence either $\tau$ or $\Delta^2_{21}(k)$ (i.e. confounding variables), and standard atomic-cooling and minihalo (Pop~III.2) parameters, for a total of 18 free parameters.
Priors are tabulated in \autoref{tab:fisher_params}, with conservative Gaussian priors.
For the Pop~III.1 source parameters, these come from the literature \citep{Sanati2025, Petkova2026}, with widths large enough to fit literature adopted ranges to within $1-1.5\sigma$.
The priors on $f_{\ast,10}$ and $\alpha_\ast$ come from prior fits to UV luminosity function data with 21cmFAST \citep{Park2019}, and the priors on $f_{\rm esc,10}$ and $f_{\ast,7}^{\rm mini}$ are physically constrained such that they reach unity at roughly 1-2$\sigma$.
For clarity, in \autoref{tab:fisher_params} we separate the Pop~III.1 parameters into ones that directly govern the size and placement of the Pop~III.1 HII bubbles, relative to ``nuisance'' parameters that change how those bubbles impact the measured $\tau$ or $\delta T_b$ field.

Fiducial parameters were calibrated to yield $\tau\sim0.095$ and a peak $x_{\rm HII}$ redshift at $z\sim23$, and a $\tau=0.06$ for the no-Pop~III.1 case.
Note that no previous works have directly studied LW feedback on Pop~III.1 sources ($A_{\rm LW}^{\rm III.1}$), so we give it a relatively small value and set a wide prior.
Also, $C_{\rm HII}$ is specifically for modeling recombinations within Pop~III.1 relic regions.
In the normal galaxy-driven reionization we use 21cmFAST's inhomogenous recombination prescription \citep{Sobacchi2014}.
Priors on $f_{\rm esc,10}$ and $f_{\ast,10}$, for example, set maximum values of unity at $2\sigma$ probability.

Assuming a Gaussian likelihood, the elements of the posterior Fisher matrix \citep{Fisher1925, TegmarkTaylorHeavens1997, CoultonWandelt2023} are computed as,
\begin{align}
\label{eq:fisher}
F_{ij} = \frac{\partial {\bf d}^T}{\partial\theta_i}{\bf C}^{-1}\frac{\partial{\bf d}}{\partial\theta_j} + F^{\rm prior}_{ij},
\end{align}
where $\bf d$ is the data vector holding the 21\,cm power spectra and (optionally) the 21\,cm global signal, and $\bf C$ is the covariance of the data vector.
$F_{ij}^{\rm prior}$ is the contribution to the posterior Fisher matrix from the prior, ${\bf F}_{\rm prior} = {\bf C}_{\rm prior}^{-1}$, where ${\bf C}_{\rm prior}$ is a diagonal covariance matrix holding the prior variances from \autoref{tab:fisher_params}.
The data covariance is assumed diagonal, with a variance on the spherically-averaged power spectrum coming from the combined thermal noise and sample variance error budget \citep{Murray2024}.
The sample variance calculation uses a fiducial 21\,cm power spectrum and the cosmological volume probed by the total survey area (i.e. transverse area) and power spectrum bandwidth (i.e. line of sight depth).
To compute the marginalized uncertainty on the parameters given the likelihood and prior we perform a dense inversion of \autoref{eq:fisher} and report the diagonals,
\begin{align}
\sigma_i^2 = [{\bf F}^{-1}]_{ii}.
\end{align}
To compute the uncertainty on derived parameters that are not explicit knobs in our simulation (e.g. $\tau$) we can project the posterior covariance with the parameter Jacobian,
\begin{align}
\label{eq:sigma_tau}
\sigma_\tau^2 = \nabla\tau^T{\bf F}^{-1}\nabla\tau,
\end{align}
where $\nabla\tau$ is the Jacobian $\partial\tau/\partial\theta$.
In all cases considered the Fisher matrices (\autoref{eq:fisher}) are well-conditioned.

\begin{figure*}
\centering
\includegraphics[width=.49\linewidth]{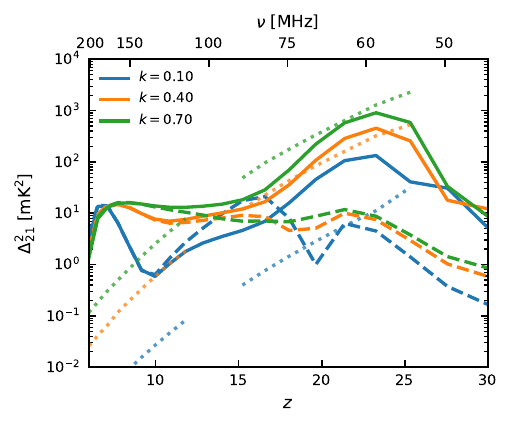}
\includegraphics[width=.49\linewidth]{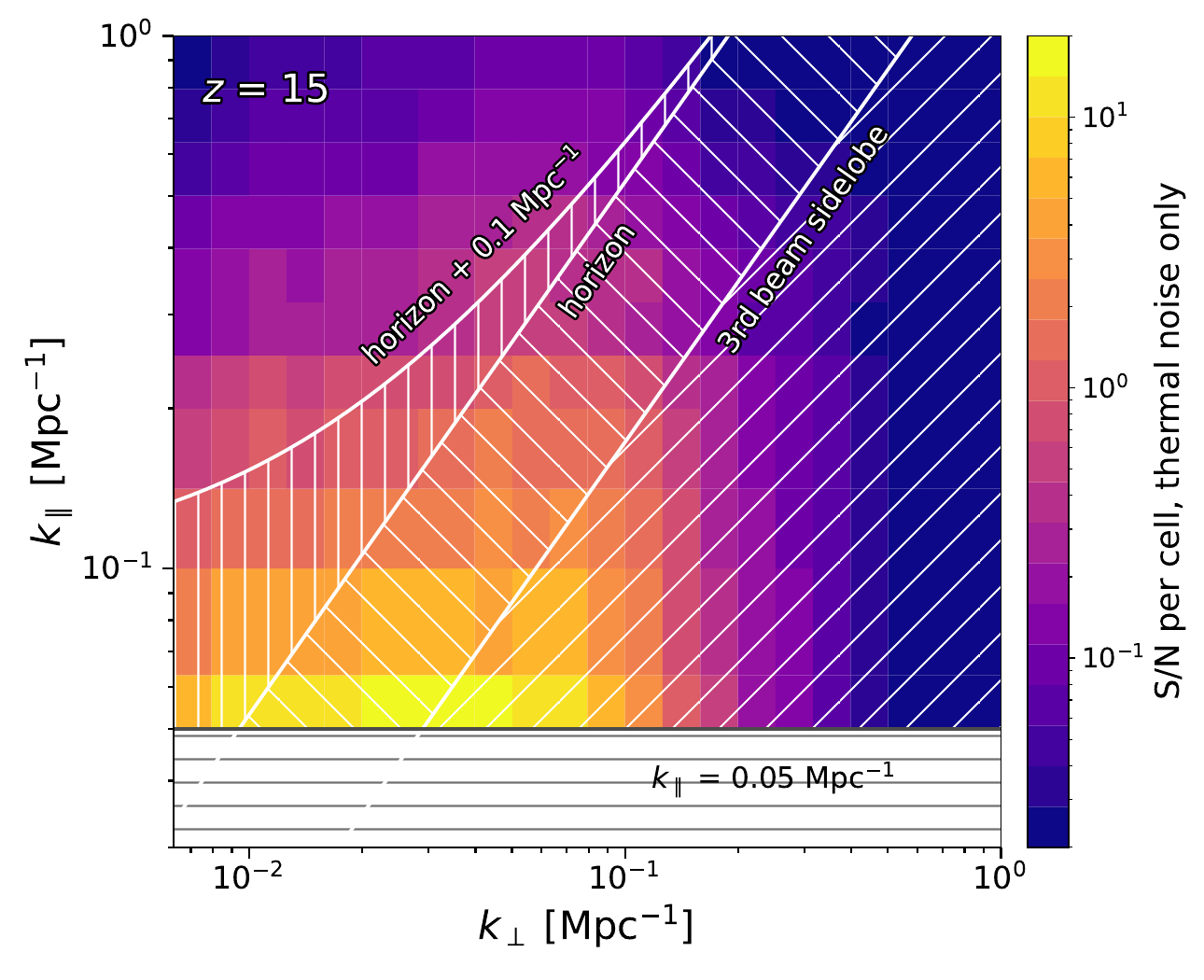}
\caption{
{\bfseries Left:} Fiducial 21\,cm power spectrm against redshift for three accessible wavenumbers with (solid) and without (dashed) Pop~III.1 physics.
We plot a fiducial SKA-Low AA4 $1\sigma$ thermal noise amplitude after 1080 hours of multi-field observations (dotted) with foreground subtraction.
The noise curves omit data for $\nu<55$ MHz where SKA has little sensitivity, and in the $90-110$ MHz range where FM radio is typically a heavy contaminant, regardless of location on Earth.
{\bfseries Right:}
Signal-to-noise of the fiducial 21\,cm power spectrum in cyclindrical $(k_\perp,k_\parallel)$ space at $z=15.2$. We show the four adopted foreground treatment strategies studied in this work: subtraction (``fg-sub'', horizontal hatch), avoidance to the beam's 3rd sidelobe (``SL''), avoidance to the horizon (``H''), and avoidance to the horizon plus a constant buffer (``H+B''). 
In each case we truncate data within the hatched region before forming our likelihood.
Thus, ``fg-sub'' preserves the most data (and is the most optimstic) while ``horizon+buffer'' discards the most data (and is the most pessimistic).
Note that even our optimistic foreground subtraction scenario does not imply perfect foreground removal: we always truncate data for $k_\parallel<0.05$ Mpc$^{-1}$ in all scenarios.
}
\label{fig:ps_noise}
\end{figure*}

To approximate the gradients, $\partial{\bf d}/\partial\theta=\partial\Delta^2_{21}/\partial\theta$, numerically, we use a central finite difference, taking steps in the positive and negative direction centered on our fiducial parameters (\autoref{tab:fisher_params}).
These steps are generally 5-10\% of the fiducial parameter values, and have been manually tuned to yield a stable gradient response.
Furthermore, our use of a central difference is 2nd order accurate over a single step forward difference.
For each of these parameter gradients we run four independent simulations with a different random number generator (RNG) for the dark matter initial conditions.
To make the most of these runs we use the de-biased Fisher estimator from \citet{CoultonWandelt2023}.
With this Fisher estimator, we find that four independent random seeds is enough to reduce the numerical noise of the simulation to below the data covariance, finding convergence in the derived parameter constraints when using 2, 3, and 4 of the random seeds.

In this section we quote two main figures of merit: the marginalized 1$\sigma$ uncertainty on the optical depth parameter, $\sigma_\tau$, and the ``detection significance.''
The former is computed from \autoref{eq:sigma_tau}, the former is computed by a likelihood ratio test between the fiducial Pop~III.1 simulation and the same simulation with Pop~III.1 physics turned off.
This asks to what significance could we detect the presence of Pop~III.1 signatures in our data relative to a no-Pop~III.1 scenario based on the noise levels of the adopted 21\,cm experiments.
This is computed only for redshifts affected by Pop~III.1 physics ($z>15$), does not marginalize over standard galaxy parameters in 21cmFAST, and can therefore be thought of as a rough heuristic of how much new signal in the data is introduced by Pop~III.1 physics, weighted by the 21\,cm survey variance.

\begin{figure}[!t]
\centering
\includegraphics[width=\linewidth]{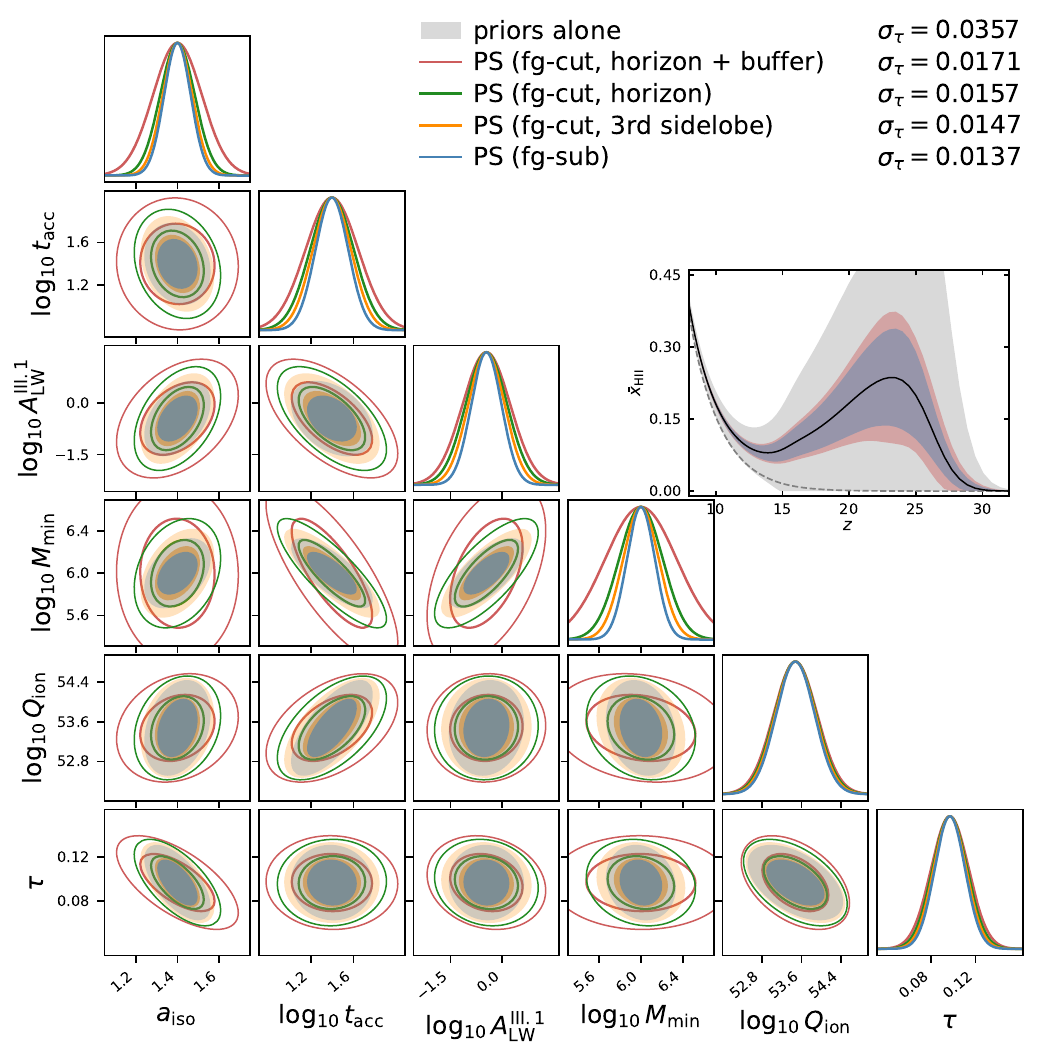}
\caption{
Corner plot of five Pop~III.1 parameters with a $\Delta^2_{21}(k)$-only power spectrum likelihood, evaluated only for $z>=15$ and varying only the ten Pop~III.1 parameters (\autoref{tab:fisher_params}).
We plot 1 and 2$\sigma$ contours for four different foreground cuts in the power spectrum described in \autoref{fig:ps_noise}. The prior is sufficiently broad to encapsulate a no Pop~III.1 flash scenario on the global ionization history $x_{\rm HII}$ within $1\sigma$ (grey shaded).
In an optimistic foreground scenario (fg-sub), the 21\,cm power spectrum alone can measure a non-zero $x_{\rm HII}$ at $z\sim23$ in our fiducial simulation. Although we only show six parameters, the posterior is computed by marginalizing over all ten Pop~III.1 parameters described in the text. 
}
\label{fig:corner_ps}
\end{figure}

\subsection{Noise profiles}

We use the 21cmSense package \citep{Pober2014, Murray2024} to forecast the power spectrum thermal noise profile of SKA Low in its AA4 configuration \citep{Braun2019}.
We assume 1080 hours of coherent observations in a multi-field survey strategy, surveying a total of $12\ {\rm deg}^2$ at $z=6$ and $170\ {\rm deg}^2$ at $z=25$.
We adopt a sky temperature model of $T_{\rm sky}=260\ {\rm K}\ (\nu/150\ {\rm MHz})^{-2.6}$, and a receiver temperature of 100 K.
Our power spectra are estimated across 20 redshift bins between $6 < z < 25$ (200 -- 55 MHz), with a bandwidth that holds $\Delta a/a\sim0.08$ roughly constant ($\Delta\nu=14$ MHz at $z=6$ and $\Delta\nu=4$ MHz at $z=25$).
In 21cmSense, for each wavemode probed by the powerspectrum, the thermal noise root mean square (RMS) is computed as,
\begin{align}
\label{eq:PN}
\Delta^2_{N}(k) = X^2Y\frac{k^3}{4\pi^2t_{\rm int}}\Omega_{\rm eff}T_{\rm sys}^2\ [{\rm mK}^2],
\end{align}
where $T_{\rm sys}=T_{\rm sky}+T_{\rm receiver}$, $t_{\rm int}$ is the total coherent integration time for a given $k$ mode, $X^2Y$ converts angles and frequencies into comoving volume, $X=D_c(z)$, $Y=c(1+z)^2/[H(z)\nu_{21}]$, and $\Omega_{\rm eff}$ is the effective SKA station beam area in steradians \citep{Parsons2012, Pober2014}.

We exclude bins at $z>25$ where the SKA observing band cuts off, but we note that extensions to these higher redshifts could better constrain the onset of the Pop~III.1 stage. 
We also take the conservative approach of truncating power spectra in the frequency range $90 < \nu < 110$ MHz ($15 > z > 12$), which are typically contaminated by terrestrial FM radio frequency interference (RFI) even in radio quiet zones. 
\autoref{fig:ps_noise} (left) shows these thermal noise amplitudes (dotted) for our SKA-Low survey specifications, plotted against our fiducial Pop~III.1 run (solid).
We also use 21cmSense to compute the cosmic variance errorbar given our observing volume and a fiducial power spectrum model (not plotted), which are added to the thermal noise error to form the total covariance in our Fisher matrix forecast \citep{Pober2014}.

\subsection{Foreground modeling}

Smooth-spectrum foregrounds dominate the 21\,cm signal by orders of magnitude. 
These are generally dominated by synchrotron and free-free emission from the galaxy and extragalactic radio sources. 
Considerable work has gone into modeling these radio foregrounds \citep[e.g.][]{Jelic2008, Bernardi2013, Remazeilles2015, Zheng2017, HurleyWalker2017, Mozdzen2019, Line2020, Gehlot2022}.
Additionally, there is considerable work understanding the instrumental transfer function that introduces spectral chromaticity on top of these smooth-spectrum foregrounds, and how this limits our ability to robustly extract the 21\,cm signal.
21\,cm foreground and instrumental modeling and removal is a highly active area of study, and is where most of the signal gains will come in the near term as most 21\,cm experiments are either foreground or instrumental systematics limited.

For a forecasting exercise, however, it is useful to consider a few broad scenarios for how well foreground and other systematics could be suppressed in future 21\,cm datasets.
Foreground and instrumental effects generally lead to contamination of the 2D cylindrical power spectrum, $\Delta^2(k_\perp, k_\parallel)$, in a wedge-like feature \citep{Datta2010, Morales2012, Parsons2012, Vedantham2012, Thyagarajan2013, Pober2014}.
This can be written as a relationship between the transverse and line-of-sight wavemode,
\begin{align}
k_\parallel \le \frac{X}{\bar{\nu} Y}\sin\theta k_\perp,
\end{align}
where the inequality denotes the region in cylindrical $\Delta^2_{21}(k_\perp,k_\parallel)$ contaminated by foregrounds.
Here, $\theta$ is the angle from the telescope pointing center that is assumed to contain contaminating foregrounds, and $\bar{\nu}$ is the average observing frequency of the power spectrum bin.
Here, we study four scenarios ranging from fairly optimistic to somewhat conservative (although still pragmatic) in their assumptions of residual foregrounds.
In each case, we simply truncate the data in various regions in the cylindrical 2D.
\autoref{fig:ps_noise} (right) shows the SKA 1080 hour signal-to-noise based on our fiducial signal model in $\Delta^2(k_\perp, k_\parallel)$, showing the four scenarios for foreground treatment.

\emph{Foreground subtraction} (fg-sub): here we assume that we can model and subtract foreground (FG) and instrumental effects to high accuracy. This is the most optimisic scenario and involves removing all modes for $k_\parallel<0.05\ {\rm Mpc}^{-1}$.
We choose to truncate these modes even in a foreground subtraction case because even in an optimistic FG removal scenario, these modes are highly degenerate with the intrisnic foreground spectrum and thus difficult to recover (note this is more pessimistic than most 21\,cm forecasts that adopt a foreground removal scenario).
\emph{Foreground removal to the 3rd beam sidelobe} (3rd SL): here we assume that we can model and subtract a portion of the FG in the main field of view, where the instrument is typically well-understood and we have high SNR to model foregrounds ($\theta\sim11^\circ$ of the telescope pointing at $\nu=150$ MHz).
\emph{Foreground avoidance to the observer's horizon} (horizon): here we assume that all foregrounds are contained within the nominal observer's horizon wedge ($\theta=90^\circ$), which we discard from the data before estimating the Fisher matrix. 
\emph{Foreground avoidance to the horizon plus a buffer} (horizion + buffer): here we assume that instrumental systematics corrupt the data beyond the nominal observer horizon wedge, adding a constant $k_\parallel=0.1$ Mpc$^{-1}$ to the horizon wedge.

\subsection{Constraints with the power spectrum}

First, we perform a Fisher parameter forecast using only power spectrum data for $z>15$, marginalizing over the ten Pop~III.1 parameters in \autoref{tab:fisher_params} (fixing the galaxy parameters).
We do this to ask the question, how can 21\,cm uniquely constrain Pop~III.1 sources and their contribution to $\tau$ at high redshifts?
Many previous works have already addressed how 21\,cm can constrain $\tau(z)$ at lower redshifts $z<12$ \citep{Liu2016, Sailer2022, Shmueli2023}.
We plot the 1$\sigma$ and 2$\sigma$ posterior contours on a subset of the major Pop~III.1 parameters in \autoref{fig:corner_ps}, with separate contours for each foreground modeling scenario.
In the our optimistic scenario (fg-sub) we find a $\sigma_\tau=0.0137$, and in our more pessimistic scenario this is degraded to $\sigma_\tau=0.0171$.
As a control, we compute the constraints on $\tau$ with no data (``priors alone''), finding $\sigma_\tau\sim0.036$ and a 1$\sigma$ error on the ionized fraction history $\bar{x}_{\rm HII}$ that reaches the fiducial no-Pop~III.1 case (gray shaded).
Our detection significance for fg-sub, fg-cut (3rd sidelobe), fg-cut (horizon) and fg-cut (horizon+buffer) is 48$\sigma$, 37$\sigma$, 19$\sigma$, and 4.8$\sigma$, respectively.

The 21\,cm power spectrum measures the clustering of the relic HII regions, constraining parameters that set the size and location of sources ($a_{\rm iso}$, $M_{\rm min}$, $Q_{\rm ion}$).
This is modulated by the overall temperature contrast between the hot relics and the cold IGM, which is what leads to the significantly boosted power spectrum amplitude relative to the no-Pop~III.1 scenario at $z\sim23$ (\autoref{fig:sweep}), which we describe in more detail in the discussion below.
In \autoref{tab:forecast_allz} we tabulate $\sigma_\tau$ and detection signficance for the power spectrum across all redshifts ($6<z<25$) while marginalizing over all 18 parameters in \autoref{tab:fisher_params}.

\begin{figure}[!t]
\centering
\includegraphics[width=\columnwidth]{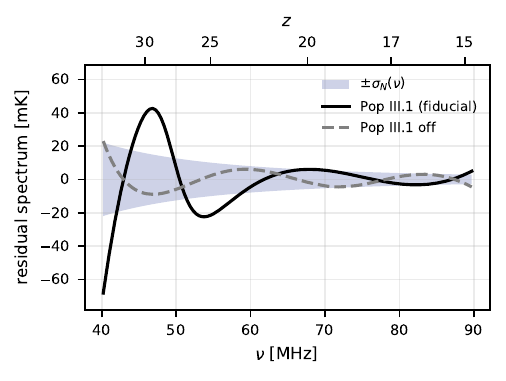}
\caption{Residuals of the global 21-cm signal over 40--90\,MHz (top axis: redshift) after fitting and removing a five-term foreground + instrument polynomial model for the fiducial Pop~III.1 model (black) and a no-Pop~III.1 model (grey, dashed).
While many global signal experiments observe down to $\nu\sim50$ MHz, we found it necessary to measure down to $\nu\sim40$ MHz to retain more information content in the global signal response to Pop~III.1 sources (with our fiducial parameters). The noise model $\sigma_N(\nu)$ is taken from the published REACH target \citep{deLeraAcedo2022}.
}
\label{fig:gs}
\end{figure}

\begin{figure}[!t]
\centering
\includegraphics[width=\linewidth]{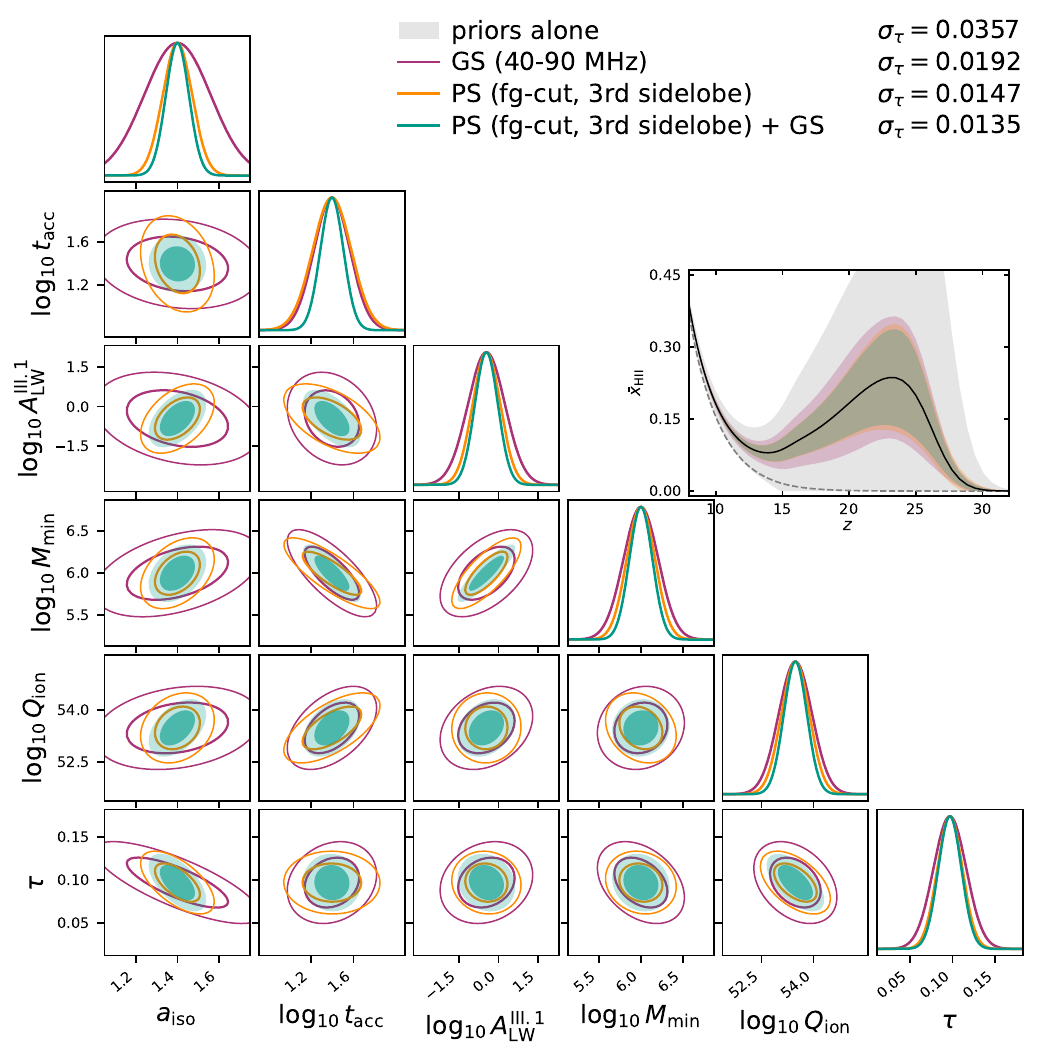}
\caption{
Combined $\Delta^2_{21}$ and global signal forecast on Pop~III.1 parameters and the optical depth.
Similar to \autoref{fig:corner_ps}, we only use 21\,cm data for $z\ge15$ and only marginalize over the ten Pop~III.1 parameters in \autoref{tab:fisher_params}.
We plot constraints from the global signal alone (GS; pink), the power spectrum alone with a 3rd beam sidelobe foreground cut (orange), and their combined likelihood (green).
The power spectrum carries most of the information content, but adding the global signal can improve sensitivity on $\sigma_\tau$ by $\sim10\%$ (for the modestly optimistic 3rd beam sidelobe foreground scenario).
The global signal, being a spatial average, discards clustering information (e.g. $a_{\rm iso}$) and is mostly sensitive to the timing of when sources turn on and when the couple the gas via Lyman-$\alpha$ production (e.g. $t_{\rm acc}$, $M_{\rm min}$).
}
\label{fig:corner_comb}
\end{figure}

\begin{figure*}[!t]
\centering
\includegraphics[width=0.7\textwidth]{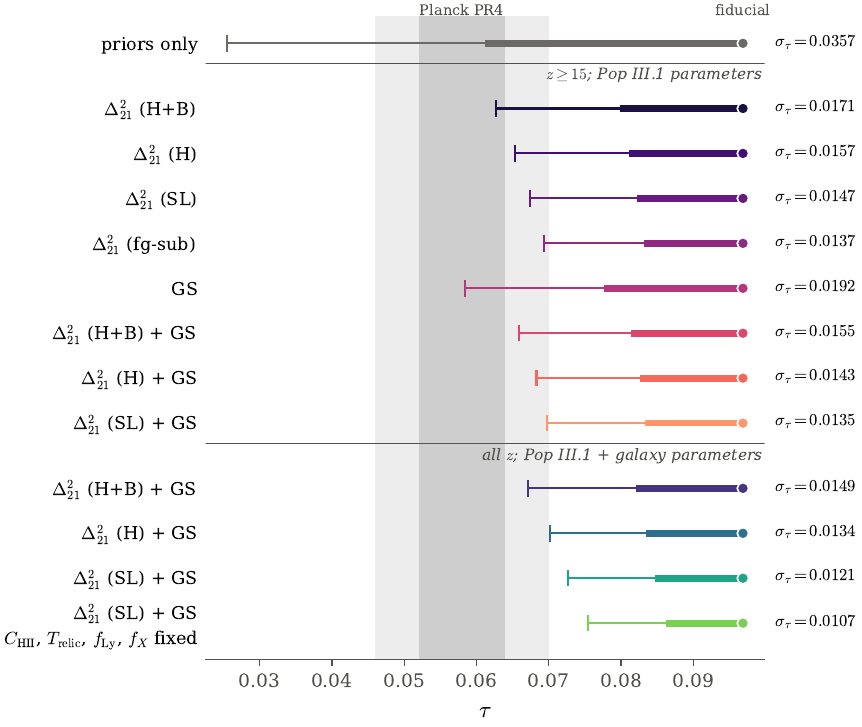}
\caption{
Forecasted uncertainty on the optical depth from 21\,cm observables.
We plot the fiducial Planck2018 constraints (grey shaded) alongside the 1$\sigma$ and 2$\sigma$ uncertainty on $\tau$ from our 21\,cm forecasts, centered on our fiducial $\tau=0.097$ simulation. 
We plot the forecasted $\sigma_\tau$ with high-redshift only data holding galaxy parameters fixed (``$z\ge15$, Pop~III.1 parameters''), in addition to a full $6<z<25$ forecast with all parameters free (``all $z$; Pop~III.1 + galaxy parameters'').
A 21\,cm interferometric (with moderate amounts of foreground removal) and global signal measurement could constrain $\tau$ to an uncertainty that would distinguish it from the Planck2018 value at $>3\sigma$ ($\Delta^2_{21}$(SL)+GS).
The foreground treatment scenarios are ordered with increasing optimisim as horizon + buffer (H+B), horizon (H), 3rd beam sidelobe (SL), and foreground subtraction (fg-sub).
Compared to when we hold the Pop~III.1 nuisance parameters fixed (``$C_{\rm HII},\, T_{\rm relic},\, f_{\rm Ly},\, f_X$ fixed''), we see an expected degredation in constraining power.
}
\label{fig:tau_summary}
\end{figure*}

\subsection{Adding the global signal}

Next we include a mock 21\,cm global signal experiment, such as EDGES \citep{Bowman2018} or REACH \citep{deLeraAcedo2022}.
We assume that spectral fluctuations from foregrounds and the instrument can be modeled and subtracted-off with a 5-term polynomial, similar to that used by EDGES \citep{Bowman2018}, which is a somewhat optimisic assumption.
Importantly, we also adopt an observing band from 40--90 MHz, different from the EDGES 50--100 MHz observing band, because we found that covering the 40--50 MHz regime was needed to maintain constraining power from the FG-filtered global signal template.
This is an assumption that is currently not planned for existing 21\,cm global signal telescopes, but could be easily adopted for future global signal efforts.

We assume that the residual noise on the foreground-subtracted spectra achieve $\sim12$ mK RMS at $\nu=50$ MHz with 0.39 MHz channelization, which is consistent with the published REACH fiducial target of 25 mK at 50 MHz with 100 kHz channelization \citep{deLeraAcedo2022}.
We scale the noise as a function of frequency as $\sigma_N(\nu)=12\,{\rm K}\, (\nu/50\,{\rm MHz})^{-2.55}$.
\autoref{fig:gs} shows the noise-free 21\,cm global signal after foreground subtraction, showing the Pop~III.1 case and the no-Pop~III.1 case against the target residual noise floor.

The noise in the raw spectrum is assumed uncorrelated, however, the foreground filtering introduces off-diagonal contributions to the noise covariance. 
If ${\bf N}_{\rm GS}$ is the diagonal matrix holding $\sigma_N^2(\nu)$ along its diagonal, then the covariance of the foreground filtered data used in the Fisher matrix is ${\bf C}_{\rm GS}={\bf N}-{\bf A}({\bf A}^T{\bf N}^{-1}{\bf A})^{-1}{\bf A}^T$, where ${\bf A}$ is the 5-term polynomial foreground basis.
\autoref{fig:corner_comb} shows the posterior parameter constraints with the combined power spectrum and global signal data vector.
For clarity, we only plot the fg-cut (3rd beam sidelobe) foreground scenario for the power spectrum.
The global signal mostly constrains timing information (when sources turned-on) through the absorption trough peak.
It weakly constrains clustering information because the spatial average discards that information.
Thus, it is more sensitive to parameters that control the timing of Lyman-$\alpha$ coupling (e.g. $t_{\rm acc}$, $M_{\rm min}$, $f_{\rm Ly}$), which \autoref{fig:corner_comb} shows can complement PS-only constraints to improve constraints on $x_{\rm HII}$ and $\tau$.

Next we forecast constraints using data at all redshifts ($6 < z < 25$) while freeing all 18 parameters in \autoref{tab:fisher_params}.
This is found in \autoref{tab:forecast_allz}.
This helps us understand to what degree Pop~III.1 signatures are degenerate with standard galaxy and minihalo physics, as prescribed by 21cmFAST.
However, at lower redshifts we also have more observational data to constrain galaxy-driven reionization, in particular Lyman-$\alpha$ forest and UV luminosity function data \citep{Greig2015, Park2019}.
Some of the galaxy parameters are poorly constrained by 21\,cm data alone, for example the degeneracy between $f_{\ast,10}-f_{\rm esc,10}$ that can by broken with a prior from fits to UVLF data \citep{Park2019}; however, this turns out to not have a strong impact on our ability to constrain $\tau$.
Ultimately, we did not incorporate a Lyman-$\alpha$ forest likelihood to constrain galaxy parameters, finding 21\,cm data alone to be sufficient in constraining the ionization fraction at $z<10$ \citep{Liu2016}: we did a test incorporating a Lyman-$\alpha$ likelihood and found no noticeable change in $\sigma_\tau$ in our PS(fg-cut, 3rd beam sidelobe) + GS scenario.
It may be the case that Lyman-$\alpha$ data does help with more pessimistic foreground cuts, which we defer to future work.

We plot the forecasted $\sigma_\tau$ for our various power spectrum and global signal datasets in \autoref{fig:tau_summary}.
We show the nominal Planck2018 preference for $\tau=0.054$ (grey shaded), and then show our 1$\sigma$ and 2$\sigma$ constraints on a Pop~III.1 driven $\tau=0.097$ scenario.
For the all-$z$, all-parameters case (bottom rows) we find $\sigma_\tau=0.0121$, which would allow for a $>3\sigma$ discrimination from the Planck2018 value in our fiducial case.
In the final row, we also show the impact of holding the Pop~III.1 nuisance parameters fixed (parameters that either change $\Delta^2(k)$ but not $\tau$ or vice versa), showing small but non-negligible degredation in $\sigma_\tau$ in our fully marginalized fit relative to holding them fixed.

\begin{table}[!t]
\caption{Forecasted optical depth uncertainty and Pop~III.1 detection significance for 21\,cm power spectrum measurements across the full $6<z<25$ range while marginalizing over all 18 parameters. For the final three rows we also include our fiducial global signal dataset.}
\label{tab:forecast_allz}
\begin{ruledtabular}
\begin{tabular}{lcc}
Scenario & $\sigma_\tau$ & Detection\\
\hline
Priors alone & 0.0399 & -- \\
\hline
fg-sub & 0.0129 & $55\sigma$ \\
fg-cut (3rd beam sidelobe) & 0.0137 & $45\sigma$ \\
fg-cut (horizon) & 0.0150 & $21\sigma$ \\
fg-cut (horizon + buffer) & 0.0166 & $8.5\sigma$ \\
fg-cut (3rd beam sidelobe) + GS & 0.0121 & $49\sigma$ \\
fg-cut (horizon) + GS & 0.0134 & $29\sigma$ \\
fg-cut (horizon + buffer) + GS & 0.0149 & $21\sigma$ \\
\end{tabular}
\end{ruledtabular}
\end{table}

\begin{figure}
\centering
\includegraphics[width=\columnwidth]{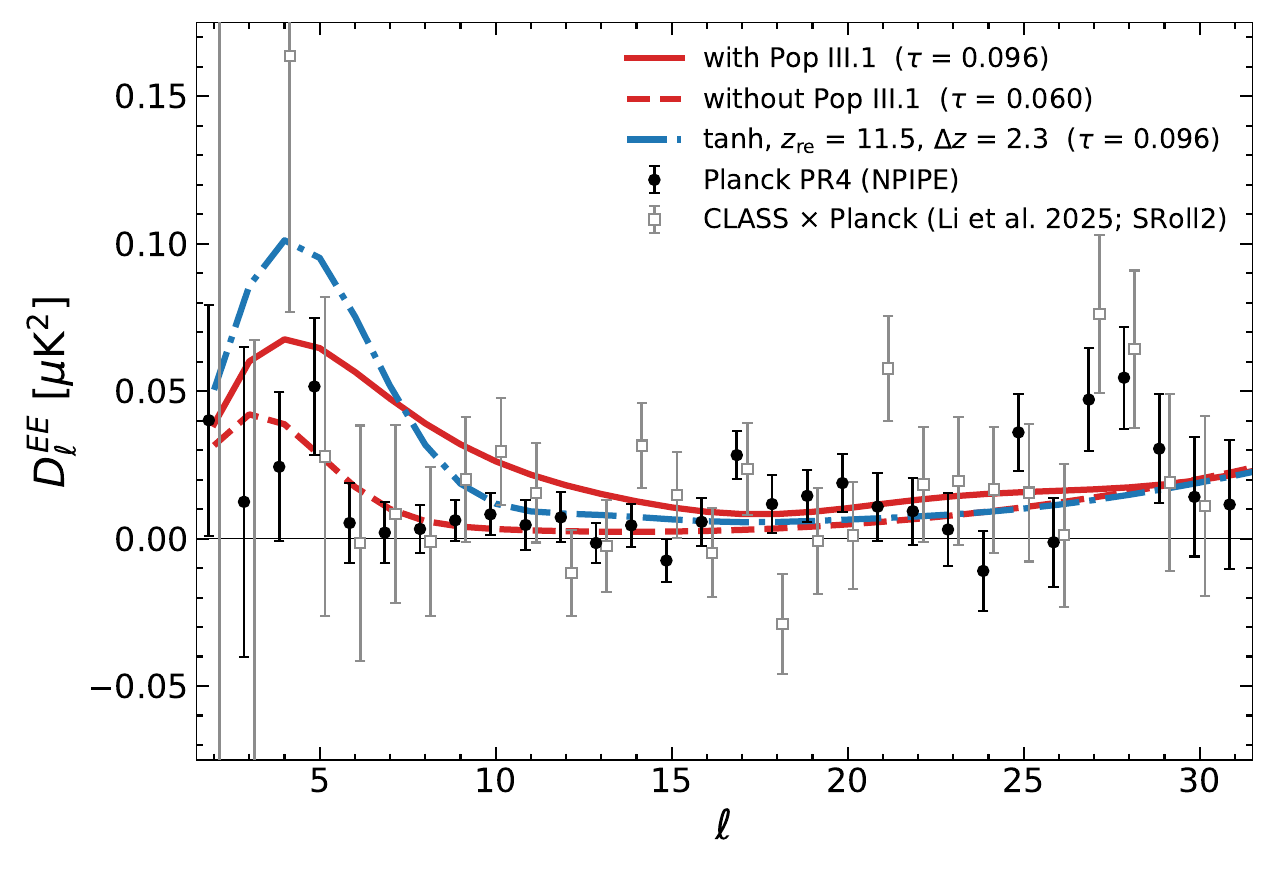}
\caption{
Low-$\ell$ CMB $E$-mode spectrum of the fiducial ionization histories, computed with the Boltzmann code
\texttt{CLASS}~\cite{Blas2011} from the mass-weighted $x_{\rm HI}(z)$ of each run. Solid red: the fiducial run with
Pop~III.1 ($\tau=0.097$). Dashed red: the same run without Pop~III.1 physics ($\tau=0.060$). Dash-dotted blue: a $\tanh$ reionization history
with the same end point ($x_{\rm HI}=0.01$ at $z=5.5$), but with a midpoint $z_{\rm re}=11.5$ and a width $\Delta z=2.3$. Black circles: the Planck PR4 (NPIPE)
$100\times143$\,GHz band powers of the LoLLiPoP likelihood~\cite{Tristram2024}, with errors from the diagonal of its
covariance. Grey squares: the cross-spectrum of the CLASS telescope's 90\,GHz map with the Planck SRoll2 100 and 143\,GHz
maps~\cite{Li2025,Delouis2019}.
The two data sets are offset in $\ell$ for clarity.}
\label{fig:lowell_ee}
\end{figure}

\begin{figure}
\centering
\includegraphics[width=8.6cm]{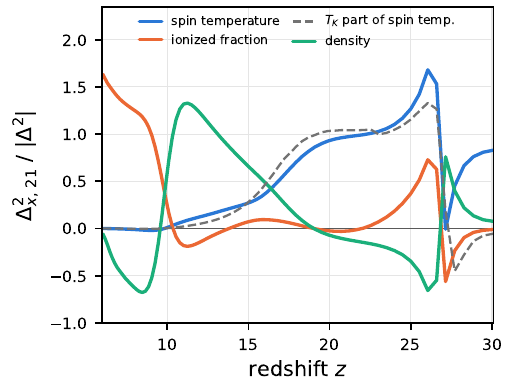}
\caption{
The fractional cross-correlation of the 21\,cm brightness temperature field with the ionization (orange), spin temperature (blue), and density field (green)at $k\sim0.2$ Mpc$^{-1}$.
This shows the relative contribution of these various terms to the observable 21\,cm power spectrum.
At late times $z<10$, the ionization drives the 21\,cm clustering signal whereas at earlier times it is driven by the density field and spin temperature fluctuations.
For the purposes of extracting $\tau$, we care about the ionization field, which at high redshifts contributes only a portion of the 21\,cm fluctuations.
Thus constraints on $\tau$ from $\Delta^2(k)$ at $z > 15$ are to first-order mediated through the model's connection between $x_{\rm HII}$ and $T_S$.
However, the spin temperature fluctuations (in our Pop~III.1 model) are driven mostly by saturated temperature fluctuations (dashed black), not by fluctuations in the Lyman-$\alpha$ coupling strength, $x_\alpha$, making it slightly less model dependent.}
\label{fig:cross_power}
\end{figure}

\section{Discussion}
\label{sec:discussion}

\subsection{CMB E-mode polarization data}

From a cosmological perspective, any model of high-redshift structure formation that can inject free electrons into the IGM could boost the optical depth to higher values of $\tau\sim0.09$ without coming into conflict with lower redshift data constraining the end of reionization \citep{Elbers2025, Weiner2026}. 
This could naturally reduce cosmological tensions with $\Lambda$CDM coming from combined probes of early universe and late-time clustering that are beginning to show a preference for anomalously low neutrino mass and dynamical dark energy \citep{DESIDR2, elbers2025constraintsneutrinophysicsdesi, Sullivan2026}.
Constraints that limit $\tau\le0.07$ come either from the low-$\ell$ CMB E-mode polarization data \citep{PlanckInt2016XLVI, Planck2020}, or from Lyman-$\alpha$ observations that also implicitly assume the IGM to be completely netural for $z > 15$ \citep[e.g.][]{Kageura2026, MonteroCamacho2026}.
Indeed, if we take the low-$\ell$ CMB E-mode data at face value it would suggest $\tau\sim0.055$; however, its robustness to polarization systematics particularly at large angles where the reionization history information lies has been questioned, even though there is no direct evidence for residual systematics in the data currently \citep{PlanckInt2016XLVI, Delouis2019, Tristram2024}.
For a standard late time, rapid reionization model (usually modeled as a tanh function in redshift), the reionization information content in the E-mode power spectrum comes predominantly from $\ell\le8$, known as the ``reionization bump.''
Ground-based experiments that only probe a fraction of the sky are therefore hard-pressed to achieve highly competitive constraints on $\tau$ at these very large angles.
A next-generation, all-sky, space-based CMB polarization experiment specifically tuned for large-angle polarization measurement such as LiteBIRD \citep{LiteBIRD2023} could be the ultimate test of this; however, the exact timeline for such a mission is still not fully clear.

As pointed out recently in \citep{TanKomatsu2025, Upadhyay2026, Weiner2026}, transient phases of IGM ionization at redshifts $z > 6$ can introduce unique signatures in the E-mode power spectrum at higher $\ell$ modes, which correspond to smaller horizon scales at earlier times.
Thus, a pathway for modern, ground-based CMB polarization experiments that target the intermediate, low-$\ell$ modes of the E-mode power spectrum such as CLASS \citep{Li2025} may be able to weigh-in on the high-redshift ($z > 12$) contributions to $\tau$ on a faster timeline.
In \autoref{fig:lowell_ee}, we plot the Planck PR4 E-mode power spectrum data (black solid) against a standard $\tau\sim0.06$ no-Pop~III.1 reionization model (red dashed) that fit those data well \citep{Tristram2024}.
To acheive a high value of $\tau$ with a standard tanh model of galaxy-driven reionization, one must boost the $\ell\sim5$ power beyond what the Planck data support (blue dashed).
However, an extended, transient reionization history like the models considered in this work leave a distinctly different imprint on the E-mode power spectra, showing a more extended distribution of power out to high $\ell$ modes.
This was pointed out recently by \citep{TanKomatsu2025}, who made similar calculations with their analytically-drived Pop~III.1 reionization history.
Here we present the first E-mode power spectrum predictions from a self-consistent Pop~III.1 source and feedback treatment in a cosmological simulation.
In \autoref{fig:lowell_ee} we also plot the recent low-$\ell$ E-mode measurements from CLASS \citep{Li2025}, shown in cross-correlation with Planck PR4 data (grey).
This data is understandably noisier than the Planck data, and at $\ell<6$ is cosmic variance limited with an $f_{\rm sky}\sim0.5$, however, it serves to visually convey that future ground-based CMB E-mode experiments may begin to probe the $\ell\ge10$ CMB E-mode power spectrum with modern polarization systematics rejection hardware, and possibly meaningfully weigh-in on high-$z$ contributions to $x_{\rm HII}(z)$ and $\tau$, independent of Planck.

\subsection{Model dependence}
When discussing the robustness of 21\,cm constraints on $x_{\rm HII}$ and thus $\tau(z>15)$, it is important to consider its model dependence.
In \autoref{fig:cross_power} we plot the fractional cross-power spectrum (with respect to the auto-power spectrum) of the 21\,cm brightness field with the ionization field, the density field, and the spin temperature field, which are all components of the observed 21\,cm signal (\autoref{eq:21cm}).
At low redshifts at the mid and end points of galaxy-driven reionization ($z < 10$) the 21\,cm fluctuation signal is driven by ionization field fluctuations.
This makes is a cleaner probe of $x_{\rm HII}$ and thus as a consequence contributions to $\tau$ at those redshifts, as has been pointed out by previous works \citep{Liu2016, Shmueli2023}.
While the 21\,cm signal still probes the ionization field at higher redshifts, its relative contribution is considerably smaller, as shown in \autoref{fig:cross_power}.
At the redshifts of interest ($z > 15$) in our fiducial simulation, the 21\,cm power spectrum $\Delta^2_{21}(k)$ is driven primarily by spin temperature fluctuations, and its connection to the ionization field is therefore model dependent.
On one hand, this might suggest that there is little hope to constrain $\tau(z>15)$ purely from 21\,cm observations in a model indenpendent manner, which may be a true statement.
However, we can take steps to make a 21\,cm-driven constraint on $\tau$ less model dependent, even in the context of using a semi-numerical simulation like 21\,cmFAST.
First, what we do in this work, is to identify some of the major nuisance parameters that act as confounding variables in the relationship between $\Delta^2_{21}$ and $\tau$. Specifically, we have identified these as the Pop~III.1 nuisance parameters of \autoref{tab:fisher_params}, $C_{\rm HII}$, $T_{\rm relic}$, $f_{\rm Ly}$, and $f_X$, which either change $\Delta^2_{21}$ without changing $\tau$ or vice versa.
By marginalizing over these paramaeters in the Fisher matrix (instead of holding them fixed) we are marginalizing-out their influence on our constraints, which does have an impact on the sensitivity of our inference at the benefit of less model dependence (\autoref{fig:tau_summary}).

A more exhaustive study of how the 21\,cm-only inference can be made less model dependent is deferred to future work.
However, one reason for optimism found in \autoref{fig:cross_power} is the fact that, in a supermassive Pop~III.1 model, most of the spin temperature fluctuations are driven by temperature fluctuations at $z<27$, as opposed to fluctuations in the Lyman-$\alpha$ WF coupling strength, $x_\alpha$.
This follows from an assumption that the gas is already strongly coupled due to efficient Lyman-$\alpha$ emission.
In particular, at the redshfits where the power spectrum peaks ($z\sim23$), the strong boost in $\Delta^2_{21}$ amplitude over fiducial models is driven largely by the contrast between hot relic bubbles and the adiabatically cold intervening IGM between them.
The hot relic gas is significantly above the CMB temperature, meaning the spin temperature component in \autoref{eq:21cm} saturates and is largely independent of the exact value of $T_{\rm relic}$, with the adiabatically cold IGM also being straightforward to compute.
Standard astrophysical sources do not produce such bubbles at $z\gtrsim20$: X-rays heat the gas smoothly, and the ionized regions of Pop~III stars in minihalos are far smaller than $\sim$Mpc scale.
Thus, a generic, effective model for the 21\,cm power spectrum at these redshifts becomes simply the geometry, size, and clustering of hot bubbles in a cold IGM, which could describe a broad class of transient reionization scenarios with few assumptions.
We leave its development to future work.
This does make some assumptions about the timing of when these sources turn on and the WF coupling strength, but these are also processes that are conveniently constrained by the global signal.

It should be re-iterated that the standard WIMP annihilation-supported supermassive Pop~III.1 scenario explored in this work is sensitive to particle physics parameters: WIMP particle mass, annihilation cross-sections, and the decay products all control how much energy is actually deposited into the surrounding medium.
To first order, changes in these parameters impact the effectiveness of energy deposition into the protostellar cloud, which ultimately sets the accretion timescale, $t_{\rm acc}$, and the mass of the object before it collapses onto the ZAMS, which impacts $Q_{\rm ion}$.
Thus, by varying these Pop~III.1 source parameters in our fit we are also effectively varying a degenerate combination of underlying particle physics parameters that are the underpining to this particular model.
Future work could explore how one might convert constraints on Pop~III.1 source parameters to constraints on dark matter properties in the context of a semi-numerical model like 21cmFAST.

\section{Conclusions}
\label{sec:conclusions}

We have presented a detailed analysis of the first self-consistent semi-numerical model of supermassive Pop~III.1 feedback on the IGM and the 21\,cm signal.
Our source model places supermassive Pop~III.1 objects (also referred to as dark stars in the literature) within collapsed halos while accounting for radiative feedback that disturb the pristine gas needed to form these objects, suppressing their formation.
We incorporate this set of physics into the existing 21cmFAST semi-numerical code, using its infrastructure to account for ionizing, Lyman-band, and X-ray radiative feedback.
We also track recombination and cooling in flash-heated relic HII regions and track the inhomogenous 21\,cm spin temperature within the IGM.
Supermassive Pop~III.1 sources leave a distinct signature in the 21\,cm global signal via early Lyman-$\alpha$ coupling that moves the absorption trough to very high redshifts that is difficult to reproduce through standard astrophysics.
They also dramatically increase the observed 21\,cm power spectrum at high redshifts ($z>20$), making its detectability more promising.

For the 21\,cm power spectrum, we show that most of its signal is driven by spin temperature fluctuations at high redshift $z>15$ in a Pop~III.1 scenario.
Thus, connecting the observed 21\,cm signal to the ionization field is somewhat model dependent, and should be interpreted with caution.
However, most of the signal boost in $\Delta_{21}^2(k)$ at $z > 20$ comes from $T_K$ fluctuations, driven by the sharp contrast between hot relic bubbles and the surrounding adiabatically cold IGM, the former of which is heated above the $T_S$ saturation point and the latter of which is straightforwardly predicted from adiabatic cooling laws.
This could allow for less model dependent constraints on the high-redshift ionization history from 21\,cm observables, even if the signals are to first order tracing spin temperature fluctuations.
We take initial steps to reduce the model dependence of our results by marginalizing over Pop~III.1 ``nuisance parameters'' (parameters that change the 21\,cm power spectrum but not $\tau$ and vice versa); however, a more detailed study of marginalization over spin temperature fluctuations is left for future work.
Our forecasts suggest that these 21\,cm observables should be detectable with upcoming 21\,cm interferometric radio surveys like the SKA, and possibly with future 21\,cm global signal experiments provided they can contain instrumental spectral structure, can cover the radio band down to 40 MHz, and can reduce foreground-subtracted residuals to $<$20 mK.
We show that future 21\,cm measurements (e.g. from the SKA) could constrain a Pop~III.1-driven contribution to the optical depth at high redshifts at roughly $\sigma_\tau\sim0.012$, with a moderately optimistic assumption regarding foreground removal.
For a fiducial optical depth of $\tau=0.097$, this would distinguish it from a Planck value of $\tau=0.055$ at $>3\sigma$, independent of low-$\ell$ CMB E-mode polarization data.
Independent of the constraints they set on $\tau$, 21\,cm measurements will be able to detect the presence of Pop~III.1 physics in the deviations they produce from standard astrophysics at high significance.

Future work should focus on tackling second-order approximations left out in this work on the modeling side.
This includes catalytic feedback on Pop~III.2 formation: currently we only model Pop~III.2 suppression via Lyman-Werner feedback; however, leftover free electrons from Pop~III.1 sources can catalyze H$_2$ cooling, possibly making Pop~III.2 formation more efficient, which we currently omit.
We also do not model redshift space distortions or light-cone effects in the power spectrum.
Another avenue for future work should be looking at complementary observables to constrain high-redshift reionization histories.
For example, in a companion paper, we study the impact that these supermassive Pop~III.1 sources have on the kinetic Sunyaev-Zeldovich (kSZ) signal, showing it to be another probe of high-redshift contributions to the optical depth \citep{Bansal2026}.
In particular, we show that our self-consistent treatment changes the kSZ signal in the supermassive Pop~III.1 scenario from what previous works have predicted, which notably allows for a $\tau\sim0.09$ scenario without coming into strong conflict with existing kSZ constraints.
Combined estimators, such as the kSZ$^2\times$21\,cm estimator may in fact be an even more model-independent way to probe the ionization field at high-redshifts, and deserves attention.
Ultimately, new probes of high-redshift ($z>12$) contributions to the hydrogen reionization history will be vital to fully understanding the interplay between the optical depth parameter and hints of new physics from modern cosmological surveys.

\begin{acknowledgements}

The authors gratefully acknowledge Simone Ferraro, Kendrick Smith, Xiangyu Jin, and Feige Wang for helpful discussions during the work.
This work has been supported by the Department of Energy under contract
DE-SC0019193.
This research was supported in part through computational resources and services provided by Advanced Research Computing at the University of Michigan, Ann Arbor.
We gratefully acknowledge support from the Leinweber Institute for Theoretical Physics at the University of Michigan.

AI tools were used to generate some of the code used in this work, in addition to outlining and proofreading some of the manuscript text.

\end{acknowledgements}

\bibliographystyle{apsrev4-2}
\bibliography{refs}
\end{document}